\documentclass[iicol,sn-mathphys-num]{sn-jnl}

\usepackage{graphicx}%
\usepackage{multirow}%
\usepackage{amsmath,amssymb,amsfonts}%
\usepackage{amsthm}%
\usepackage{mathrsfs}%
\usepackage[title]{appendix}%
\usepackage{xcolor}%
\usepackage{textcomp}%
\usepackage{manyfoot}%
\usepackage{booktabs}%
\usepackage{algorithm}%
\usepackage{algorithmicx}%
\usepackage{algpseudocode}%
\usepackage{listings}%
\usepackage{float}

\theoremstyle{thmstyleone}%
\theoremstyle{thmstyletwo}%

\theoremstyle{thmstylethree}%

\begin{document}

\title[Bright coherent attosecond X-ray pulses from beam-driven relativistic mirrors]{Bright coherent attosecond X-ray pulses from beam-driven relativistic mirrors}


\author*[1,2]{\fnm{Marcel} \sur{Lamač}}\email{marcel.lamac@eli-beams.eu}

\author[1]{\fnm{Petr} \sur{Valenta}}

\author[1,3]{\fnm{Jaroslav} \sur{Nejdl}}

\author[1]{\fnm{Uddhab} \sur{Chaulagain}}

\author[1]{\fnm{Tae Moon} \sur{Jeong}}

\author[1,4]{\fnm{Sergei V.} \sur{Bulanov}}

\affil*[1]{\orgdiv{ELI Beamlines Facility}, \orgname{Extreme Light Infrastructure ERIC}, \orgaddress{\street{Za Radnicí 835}, \city{Dolní Břežany}, \postcode{25241}, \country{Czech Republic}}}

\affil[2]{\orgdiv{Faculty of Mathematics and Physics}, \orgname{Charles University}, \orgaddress{\street{Ke Karlovu 3}, \city{Prague 2}, \postcode{12116}, \country{Czech Republic}}}

\affil[3]{\orgdiv{Faculty of Nuclear Sciences and Physical Engineering}, \orgname{Czech Technical University in Prague}, \orgaddress{\street{Břehová 7}, \city{Prague 1}, \postcode{11519}, \country{Czech Republic}}}

\affil[4]{\orgdiv{Kansai Photon Science Institute}, \orgname{National Institutes for Quantum Science and Technology}, \orgaddress{\street{8-1-7 Umemidai, Kizugawa}, \city{Kyoto}, \postcode{619-0215}, \country{Japan}}}


\abstract{Bright ultrashort X-ray pulses allow scientists to observe ultrafast motion of atoms and molecules. Coherent light sources, such as the X-ray free electron laser (XFEL), enable remarkable discoveries in cell biology, protein crystallography, chemistry or materials science. However, in contrast to optical lasers, lack of X-ray mirrors demands XFELs to amplify radiation over a single pass, requiring tens or hundreds of meters long undulators to produce bright femtosecond X-ray pulses. Here, we propose a new ultrafast coherent light source based on laser reflection from a relativistic mirror driven by a relativistic charged particle beam in micrometer-scale plasma. We show that reflection of millijoule-level laser pulses from such mirrors can produce bright, coherent and bandwidth-tunable attosecond X-ray pulses with peak intensity and spectral brightness comparable to XFELs. In addition, we find that beam-driven relativistic mirrors are highly robust, with laser-induced damage threshold exceeding solid-state components by at least two orders of magnitude. Our results promise a new way for bright coherent attosecond X-ray pulse generation, suitable for unique applications in fundamental physics, biology and chemistry.}




\maketitle

\begin{figure*}[t]
\centering
\includegraphics[scale = 0.85]{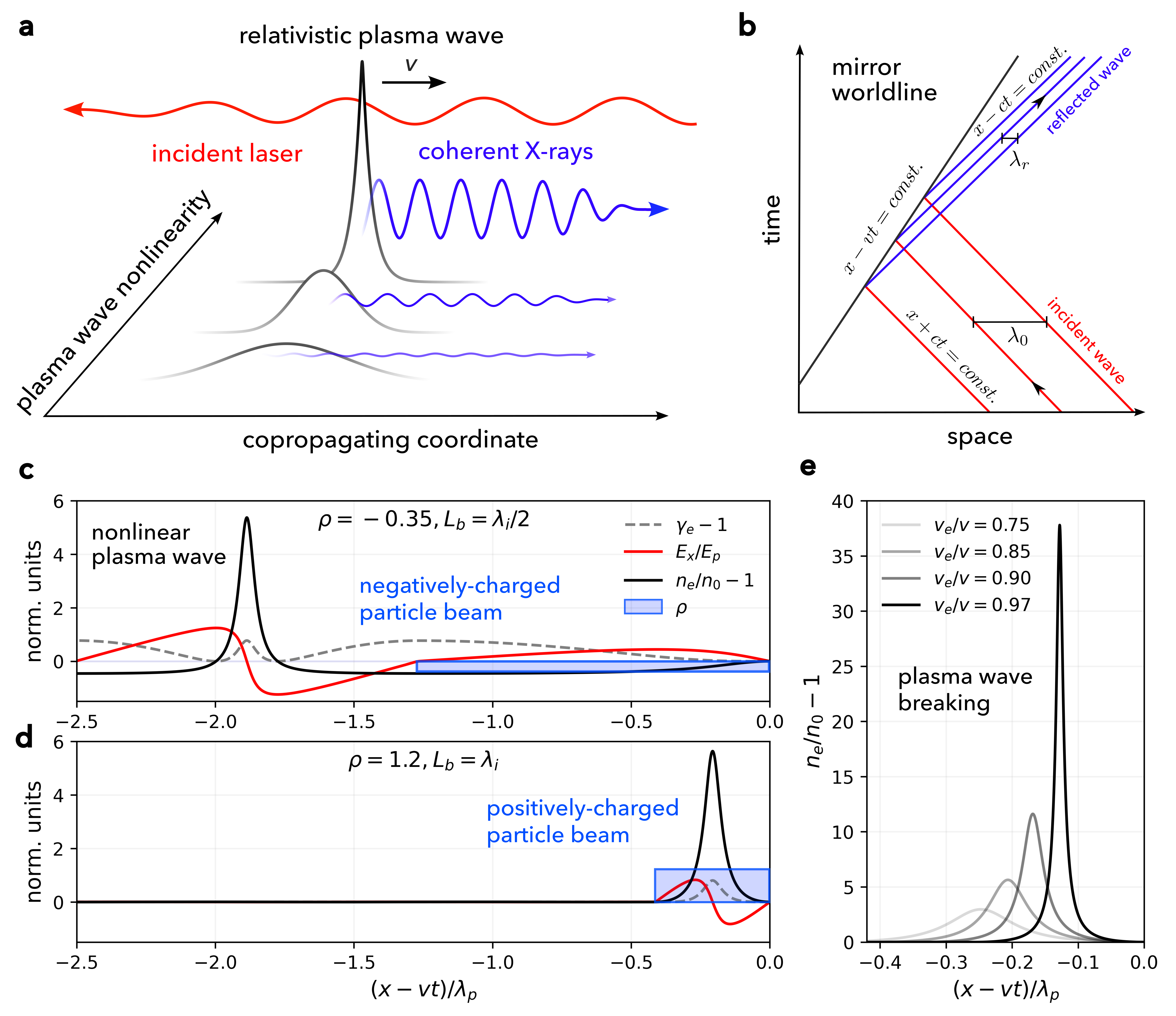}
\caption{\textbf{Relativistic mirrors driven by charged particle beams in plasma.} \textbf{a}, Schematic of a laser being reflected from a relativistic mirror formed by a propagating nonlinear plasma wave. Reflection coefficient increases with plasma wave nonlinearity, which increases the density of plasma electrons, with peak velocity $v_{e}$, which are catching up with the nonlinear wave, which propagates with velocity $v>v_{e}$. \textbf{b}, Diagram of radiation being fully reflected from a relativistic mirror propagating in spacetime with velocity $v$. At the reflection point, an electromagnetic wave undergoes a double Doppler effect, as illustrated by the wavefronts of incident (red) and reflected radiation (purple) given by constant wave-periodic values of the light-cone coordinates $x + ct$ and $x - ct$, where $x, t$ are the space and time coordinates, respectively. \textbf{c,} \textbf{d,} Numerical solutions of Eq. \ref{eq:3}, given in terms of the copropagating coordinate $x-vt$ normalized to the linear plasma wavelength $\lambda_{p}$, showing normalized electron density (black), longitudinal electric field (red) and Lorentz factor of plasma electrons (grey, dashed) driven by negatively (\textbf{c}) and positively (\textbf{d}) charged relativistic particle beams propagating in plasma with $\gamma = 5$. Normalized charge density of the driving beam $\rho$ (blue, filled) is set according to Eq. \ref{eq:4} in both cases such that electron oscillations reach $\gamma_{e} = 1.4$, producing comparable nonlinear waves. Particle beam pulse length $L_{b}$ is set according to Eq. \ref{eq:6} to drive resonant wake (negatively-charged beam, $L_{b} = \lambda_{i}/2$) and single-cycle interior (positively-charged beam, $L_{b} = \lambda_{i}$) plasma wave. \textbf{e}, Numerical solutions of Eq. \ref{eq:3}, showing that a nonlinear plasma wave, driven by a relativistic charged particle beam with $\gamma = 5$, can be brought arbitrarily close to the wave breaking threshold by tuning beam density $\rho$ according to Eq. \ref{eq:4}.}
\label{fig:1}
\end{figure*}

The theory of light reflection from a moving mirror was presented by A. Einstein in his seminal work on theory of special relativity in 1905 \citep{einstein1905elektrodynamik}. As illustrated in Fig. \ref{fig:1}b, an electromagnetic pulse reflected from a mirror counter-propagating at velocity $v$ will undergo a frequency upshift, pulse compression and amplitude increase due to a double Doppler effect, which can be written as \citep{einstein1905elektrodynamik, bulanov2003light, bulanov2013relativistic, timur2024luminal}
\begin{equation}
    \frac{\omega_{r}}{\omega_{0}} = \frac{\tau_{0}}{\tau_{r}} = \frac{E_{r}}{rE_{0}} = \frac{1+\beta}{1-\beta},\label{eq:1}
\end{equation} 
where $E_{r}, \omega_{r} ,\tau_{r}$ and $E_{0}, \omega_{0}, \tau_{0}$ are respectively the electric field amplitude, angular frequency and pulse duration of reflected and incident radiation, $\beta = v/c$ is the normalized mirror velocity, $r$ is the complex amplitude reflection coefficient (see Supplementary Information S1 for definition and derivation details) and $c$ is the speed of light in vacuum. 

This concept promises a source of coherent electromagnetic radiation with unique properties. A relativistic mirror with tunable velocity could enable measurements with fundamentally limitless spatiotemporal resolution, an ultimate prospect for many novel applications, including spectroscopy and imaging of atomic, molecular, and electronic dynamics with attosecond resolution \citep{krausz2009attosecond, neutze2000potential, o2001free, boutet2018x}, laboratory astrophysics experiments \citep{bulanov2015problems} or experimental study of nonlinear properties of quantum vacuum and other topics of fundamental interest \citep{mourou2006optics}, such as investigating black hole information paradox with accelerating relativistic mirrors \citep{chen2017accelerating}. 

Most of these applications require a coherent light source with high brightness, which is currently offered mostly by the kilometers-long and access-limited X-ray free electron lasers (XFELs) \citep{o2001free, boutet2018x}. Recently, breakthroughs in shrinking free-electron lasers were achieved by using compact plasma accelerators to produce coherent radiation in the XUV \citep{wang2021free, malaca2024coherence}, UV \citep{labat2023seeded} and infrared \citep{pompili2022free} spectral range. The next generation of coherent light sources should be compact, while ideally at the same time advancing some of the source characteristics, such as brightness, pulse duration or spectral tunability. Eq. \ref{eq:1} tells us that a robust, highly-reflective relativistic mirror with tunable velocity could satisfy all of these requirements.

The question of producing a relativisitic mirror has been a recurring topic, with reinvigorated interest today due to the availability of high-power lasers that enable its various implementations, see e.g. Refs. \citep{bulanov2013relativistic, kando2018coherent, teubner2009high} (and the references cited therein) for review, with notable examples being the relativistic mirror formed by the plasma wake wave of a laser pulse propagating in underdense plasma \citep{bulanov2003light, kando2007demonstration, pirozhkov2007frequency, kando2009enhancement, moghadasin2019attosecond, valenta2020recoil, mu2020relativistic}, the oscillating relativistic mirror driven by a laser on the surface of an overdense plasma target \citep{bulanov1994interaction, lichters1996short, vincenti2019achieving, quere2021reflecting, chopineau2022sub, lamavc2023anomalous}, or the relativistic mirror accelerated by the interaction of an intense laser with a thin solid target \citep{kulagin2007theoretical, kiefer2013relativistic, ma2014bright}. The numerous studies of laser-driven relativistic mirrors in the recent years have shown a growing maturity in the field, but none have demonstrated a robust mechanism for bright and highly-tunable coherent X-ray pulse generation. 

Relativistic mirrors formed by nonlinear plasma waves driven by lasers propagating in underdense plasma \citep{bulanov2003light, kando2007demonstration, pirozhkov2007frequency, kando2009enhancement, moghadasin2019attosecond, valenta2020recoil, mu2020relativistic} are especially attractive. The velocity of the mirror is equal to the driving laser group velocity. Changing the velocity of such a mirror experimentally therefore requires only a simple variation in gas pressure. For the usual wavelength of high-power lasers, $\lambda_{0} = 0.8\,\mathrm{\mu m}$, the electron density $n_{e}$ in a laser-driven nonlinear plasma wave can exceed the critical plasma density $n_{c}\approx 1.7\times10^{21}\, \mathrm{cm^{-3}}$, enabling coherent reflection for counter-propagating radiation as illustrated in Fig. \ref{fig:1}a. 

However, since an increase of laser group velocity equals decrease in background plasma density, the laser amplitude must be also increased to sustain sufficiently high electron density in plasma wave. This makes laser-driven mirrors sensitive to diffraction and laser instabilities, such as self-modulation or stimulated Raman scattering \citep{gibbon2005short, kruer2019physics}, which can rapidly change the laser group velocity, breaking the mirror in process and trapping the oscillating plasma electrons which sustain the nonlinear plasma wave \citep{esarey2009physics, bulanov1997transverse}. These detrimental effects make tunable generation of bright coherent attosecond X-ray pulses challenging using laser-driven relativisic mirrors.

In this work, we propose a robust and highly-tunable relativistic mirror based on a large-amplitude nonlinear plasma wave driven by a relativistic charged (both positively and negatively) particle beam propagating in plasma. We show that charged particle beams can drive a highly-reflective and stable relativistic mirrors with well-defined velocity over a relatively long distance, unlocking generation of bright and fully-coherent attosecond X-ray pulses with intensity, bandwidth and peak spectral brightness comparable to XFELs in just a few micrometers.


\section*{Results}

\subsubsection*{Relativistic mirrors driven by charged particle beams in plasma}

We begin by presenting analytical theory for beam-driven relativistic mirrors. To describe relativistic mirrors based on beam-driven nonlinear plasma waves, we consider the one-dimensional approximation, which is justified for wide relativistic particle beams satisfying $k_{p}r_{b} \gg 1$, where $k_{p} = \omega_{p}/c$ is the plasma wave number, $r_{b}$ is the radial size of the beam, $\omega_{p} = \left[e^{2}n_{0}/(m_{e}\epsilon_{0})\right]^{1/2}$ is the linear plasma frequency, $n_{0}$ is the background electron plasma density, $m_{e}$ is the electron mass, $e$ is the elementary charge and $\epsilon_{0}$ is the vacuum permittivity. The plasma waves driven by a charged particle beam can be described in terms of a co-propagating coordinate $\xi = k_{p}(x - vt)$, where $x$ is the space coordinate and $t$ represents time, by (see Supp. S2 for derivation details)
\begin{eqnarray} 
\frac{\partial^{2}\phi}{\partial \xi ^{2}} = \gamma^{2}\left[\beta\left(1 - \frac{1}{\gamma^{2}(1+\phi)^{2}}\right)^{-1/2} - 1\right] - \rho, \label{eq:3}
\end{eqnarray}
where $\phi$ is the electric potential normalized by $mc^{2}/e$, $\rho = qn_{b}/n_{0}$ is the normalized charge density of the driver, $n_{b}$ is the driver particle density and $q$ is the charge number of the driver, e.g. $q = -1$ for electrons or muons, $q = +1$ for protons or positrons and $q \geq 1$ for positively-charged ions. 

We approximate the driver in the laboratory frame with a flat-top density profile, $\rho = \mathrm{const.}$, normalized velocity $\beta$ and pulse length $L_{b}$. Plasma waves obtained by solving Eq. \ref{eq:3} numerically for such a charged (negatively and positively) driver are shown in Fig. \ref{fig:1}c, d. Highly-reflective large-amplitude electron density wave can be produced if the plasma wave is brought close to the threshold of wave breaking, where the electrons composing the plasma wave catch up with the driver, $v_{e} \rightarrow v$, or equivalently, $\gamma_{e} \rightarrow \gamma$. We solve Eq. \ref{eq:3} analytically and obtain the following maximum Lorentz factor of electrons composing the beam-driven plasma wave (see Supp. S3 for derivation details) 
\begin{equation}
    \gamma _{e} = 1 + \frac{2\rho^{2}\beta^{2}}{2\rho+1+(\rho/\gamma)^{2}},\label{eq:4}
\end{equation}
Solving Eq. \ref{eq:4} at the wave breaking threshold, $\gamma_{e} = \gamma$,  we obtain the beam charge density which causes immediate plasma wave breaking, as $\rho_{wb,+} = \gamma$ for $q > 0$ and $\rho_{wb,-} = -\gamma/(2\gamma+1)\approx -1/2$ for $q < 0$. Therefore, to drive a large-amplitude density wave over any distance, the particle beam density must be initially set below the wave breaking threshold, as $\rho / \rho_{wb} < 1$.

Large-amplitude electron density waves can be excited close to the wave breaking threshold in the wake of a charged particle beam, but if the normalized beam charge density satisfies $\rho > -1/(1+\beta) \approx -1/2$, it can be excited also in its interior as shown in Fig. \ref{fig:1}d. The following expression can be derived for the interior wave wavelength $\lambda_{i}$ in the limit $\gamma \gg 1$ (see Supp. S4 for a more general form, valid for any $\gamma$, and derivation details)
\begin{equation}
    \frac{\lambda_{i}}{\lambda_{p}} = \frac{2}{\pi}\frac{E\left[2\rho/(2\rho+1)\right]}{\sqrt{2\rho+1}}, \label{eq:6}
\end{equation}
where $\lambda_{p} = 2\pi c/\omega_{p}$ is the linear plasma wave wavelength and $E(k)$ is the complete elliptic integral of the second kind. Eq. \ref{eq:6} shows that interior waves contract (for $\rho > 0$) and expand (for $\rho < 0$) with respect to the linear plasma wave. The pulse length $L_{b}$ can be set to excite resonant nonlinear plasma wake wave when $L_{b} = (m-1/2)\lambda_{i}$, or to excite multi-cycle interior plasma waves with no wake excitation when $L_{b} = m\lambda_{i}$, where $m$ is a natural number (see Supp. S5 for derivation details). Interior plasma waves can therefore serve either as a single mirror, or a periodic train of relativistic mirrors, similarly to wake waves. 

For $\rho \leq -1/(1+\beta) \approx -1/2$, the interior wave wavelength (\ref{eq:6}) diverges and no plasma oscillations occur in the beam interior. In such a case, large-amplitude density waves can occur only in the wake of a negatively-charged particle beam which has large enough pulse length, such that plasma electrons are accelerated throughout the beam interior up to $\gamma_{e}\approx \gamma$, which can be obtained by solving Eq. \ref{eq:3} numerically.  

Beam-driven plasma waves close to the wave breaking threshold cannot propagate in plasma indefinitely. The longitudinal field of the interior plasma wave modulates the charged particle beam as it propagates. Eventually, the mirror breaks due to this relativistic beam-plasma instability. The distance $L_{wb}$ traveled by the mirror before it breaks can be obtained as (see Supp. S6 for derivation details)
\begin{equation}
    \frac{L_{wb}}{\lambda_{p}} \approx \frac{\gamma}{5}\left(\frac{M}{qm_{e}}\right)^{1/2}\left(1-\frac{\rho}{\rho_{wb}}\right)^{1/2}, \label{eq:wb}
\end{equation}
where $M$ is mass of the driver particles. Eq. \ref{eq:wb} explicitly reiterates that the mirror propagates for any significant distance only when $\rho / \rho_{wb} < 1$. In this case, we see that the mirror can be set extremely resilient, since the breaking distance grows with driver velocity and particle mass, as $L_{wb}\approx \gamma (M/m)^{1/2}$. In the following, we show that tunability of the breaking distance of beam-driven relativistic mirrors is the key that unlocks tunable generation of extremely-bright attosecond X-ray pulses.

\begin{figure*}[t]
\centering
\includegraphics[scale = 0.6]{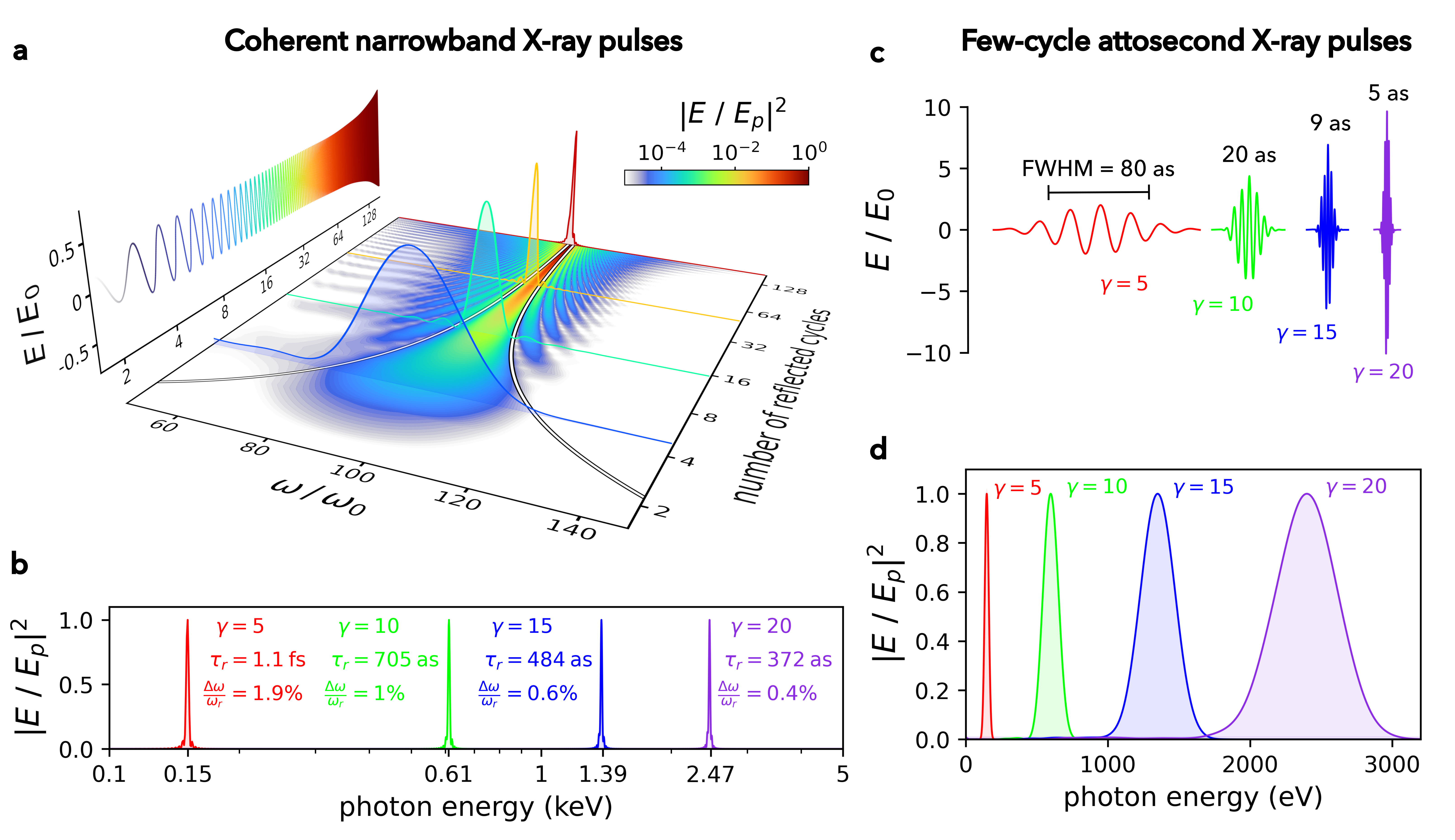}
\caption{\textbf{PIC simulation results of attosecond X-ray pulses from beam-driven relativistic plasma mirrors.} \textbf{a}, Normalized energy spectrum evolution and temporal profile of reflected radiation obtained from PIC simulations for proton-driven relativistic mirror with $\gamma = 5$ propagating in underdense plasma with $n_{0}/n_{c} = 0.01$, where $n_{c} \approx 10^{21}\text{ cm}^{-3}$ is the critical density for the incident laser wavelength $800$ nm. Theoretical bandwidth (white lines, black outline) given by $\Delta \omega/\omega_{r} = 1/N$, where $N$ is the number of reflected cycles. The inset shows temporal profile of reflected radiation corresponding to the time of spectral peak saturation. The temporal profile is colour-coded to the time (number of reflected cycles in) of corresponding normalized energy spectrum lineouts. \textbf{b}, Normalized energy spectrum of radiation reflected from relativistic mirrors driven in underdense plasma with $n_{0}/n_{c} = 0.08$ by protons with $\gamma \in \{5, 10, 15, 20\}$ and the incident laser amplitude $E_{0} = 10^{-3}m_{e}c\omega_{0}/e$. The relative spectral bandwidths $\Delta \omega/\omega_{r}$ decrease with increasing $\gamma$ due to increasing mirror propagation distance, given by Eq. \ref{eq:wb}. \textbf{c}, Temporal profile of attosecond X-ray pulses generated from beam-driven relativistic mirrors with $\gamma \in \{5, 10, 15\}$, $n_{0}/n_{c} = 0.08$ and incident laser pulse amplitude $E_{0} = 0.1m_{e}c\omega_{0}/e$. \textbf{d}, Corresponding energy spectra showing tunability from XUV to X-rays.}
\label{fig:2}
\end{figure*}

\subsubsection*{Coherent attosecond X-ray pulses}

The most important parameter of a mirror is the reflection coefficient, $r$. According to Eq. \ref{eq:1}, it determines the reflected wave amplitude, but also the number of reflected X-ray photons, $N_{\gamma,r} = |r|^{2}N_{\gamma,0}$, where $N_{\gamma ,0}$ is the number of incident laser photons. 

Reflection coefficient of beam-driven relativistic mirrors can be calculated as (see Supp. S7-S8 for derivation details)
\begin{equation}
    r \approx 0.3\left(\frac{n_{0}}{\gamma n_{c}}\right)^{2/3}\label{eq:7},
\end{equation}
where $n_{c} = m_{e}\epsilon_{0}\omega_{0}^{2}/e^{2}$ is the critical plasma density calculated for the incident laser with frequency $\omega_{0}$. While $r\rightarrow 0$ as $\gamma \rightarrow \infty$, the reflected amplitude according to Eq. \ref{eq:1} follows $E_{r}/E_{0}\approx 4\gamma^{2}r \propto \gamma^{4/3}$ due to double Doppler effect. The peak power of reflected radiation is therefore amplified as $P_{r}/P_{0} \propto \gamma^{8/3}$. Using Eq. \ref{eq:1}, we see that incident-to-reflected energy conversion efficiency also grows with mirror velocity, as $\eta = \mathcal{E}_{r}/\mathcal{E}_{0} \propto \gamma^{2/3}$, where $\mathcal{E}_{r}$ and $\mathcal{E}_{0}$ are, respectively, the total energy of reflected and incident radiation. This is valid as long as the total reflected energy does not exceed total energy of the electrons composing the relativistic mirror, $N_{\gamma,r}\hbar \omega_{r} \ll N_{e}\gamma m_{e}c^{2}$ \citep{valenta2020recoil}. In the following section, the exact threshold is formulated in terms of maximum laser fluence incident on the relativistic mirror. 


To evaluate our analytical results with self-consistent and energy-conserving numerical calculations, we have employed fully-relativistic particle-in-cell (PIC) simulations. Here, we present the results of a relativistic mirror formed by the nonlinear interior plasma wave driven by a flat-top proton beam (see Methods for simulation details), as depicted in Fig. \ref{fig:1}d, with varying mass-to-charge ratio considered in the next section. For the considered parameters, the minimum number of protons required in a beam would correspond to $N_{p} \approx \gamma n_{0}\lambda_{i}^{3} \approx 10^{7}$, which is well within the means of conventional accelerators, which can reach $N_{p}\approx 10^{11}$ \citep{muggli2017awake, verra2022controlled, caldwell2009proton, lotov2010simulation}. To investigate bandwidth tunability of reflected radiation, we consider two different cases for the incident counter-propagating laser. A low-amplitude, narrowband continuous wave (CW) and a high-amplitude, broadband few-cycle pulse. In both cases, the central wavelength is $\lambda_{0} = 0.8\:\mathrm{\mu m}$.

\textit{Continuous wave.} The spectrum evolution and the temporal profile of reflected radiation, for the case with $n_{0}/n_{c} = 0.01$ and $\gamma = 5$, are shown in Fig. \ref{fig:2}a. The CW is upshifted in accordance with Eq. \ref{eq:1}, $\omega_{r}/\omega_{0}\approx 98$. The energy spectrum peak grows quadratically in time and saturates at the the mirror breaking distance $L_{wb}$ with $N \approx 140$ reflected cycles and relative spectral bandwidth $\Delta \omega /\omega_{r} = 0.7\%$, where $\Delta\omega$ is the full-width at half-maximum (FWHM) of the reflected radiation energy spectrum. 

The energy spectrum of a reflected continuous wave can be calculated analytically (see Supp. S9 for derivation) as $|E(\omega)|^{2} = E_{p}^{2}\,\mathrm{sinc}\left((\omega-\omega_{r})\tau_{r}/2\right)$, where $E_{p}^{2} = |E_{r}|^{2}\tau_{r}^{2}$ is the spectral peak value and $\mathrm{sinc}(x) = \sin (x)/x$. The FWHM spectral bandwidth directly follows as $\Delta \omega \approx \omega_{r}/N$, where $N$ is the number of reflected cycles. This is in accordance with the PIC results, which highlights that relative spectral bandwidth is conserved upon reflection from a stable beam-driven relativistic mirror, producing a pulse with properties given exactly by Eq. \ref{eq:1}, up to the mirror breaking distance given by Eq. \ref{eq:wb}. 

Fig. \ref{fig:2}b shows normalized energy spectra saturated at the mirror breaking distance produced by proton beams, with $\gamma \in \{5, 10, 15, 20\}$ and $\rho = 0.96\rho_{wb}$, propagating in plasma with $n_{0}/n_{c} = 0.08$. While increasing background plasma density increases reflection coefficient, the number of reflected cycles drops, since the mirror breaks faster according to Eq. \ref{eq:wb}, as $L_{wb} \propto \lambda_{p}$. The fastest mirror with $\gamma = 20$ produces a fully-coherent X-ray pulse with narrow bandwidth of $10\,\mathrm{eV}$ centered at $2.47\,\mathrm{keV}$, and pulse duration $\tau_{r} \approx 372\,\mathrm{as}$ in accordance with Eq. \ref{eq:1}. 

\textit{Few-cycle pulses.} To investigate the possibility of bright few-cycle X-ray pulse generation, we set a counter-propagating few-cycle pulse with Gaussian profile incident on the nonlinear wave with peak amplitude increased to $a_{0} = 0.1$ and FWHM pulse duration $\tau_{\mathrm{FWHM}} = 3T_{0}\approx 8\: \mathrm{fs}$. Fig. \ref{fig:2}c shows the temporal profiles of radiation reflected from relativistic mirrors with three different velocities. The fastest considered mirror, with $\gamma = 20$, reflects a bright coherent X-ray pulse, as short as $\tau_{r} = 5$ attoseconds, with energy centered around $2.4\,\mathrm{keV}$ and peak electric field amplitude $E_{r} \approx 4\text{ TV/m}$, corresponding to intensity of $I_{r} \approx 2.3\times 10^{18}\text{ W/cm}^{2}$, which is of the same order as XFELs \citep{o2001free}. Fig. \ref{fig:2}d shows the corresponding normalized energy spectra. By changing the $\gamma$ of the driver, the bright few-cycle pulses are tuned according to Eq. \ref{eq:1} from XUV to X-rays. 

 Such coherent attosecond X-ray pulses could serve as an extremely bright source for unique applications which require high brightness and short pulse durations at the same time, such as ultrafast coherent X-ray spectroscopy \citep{ju2019coherent}, nonlinear XUV spectroscopy \citep{bencivenga2015four, fidler2019nonlinear} or coherent diffractive imaging of biomolecules \citep{chapman2011femtosecond}. 


\begin{figure*}[t]
\centering
\includegraphics[scale = 0.75]{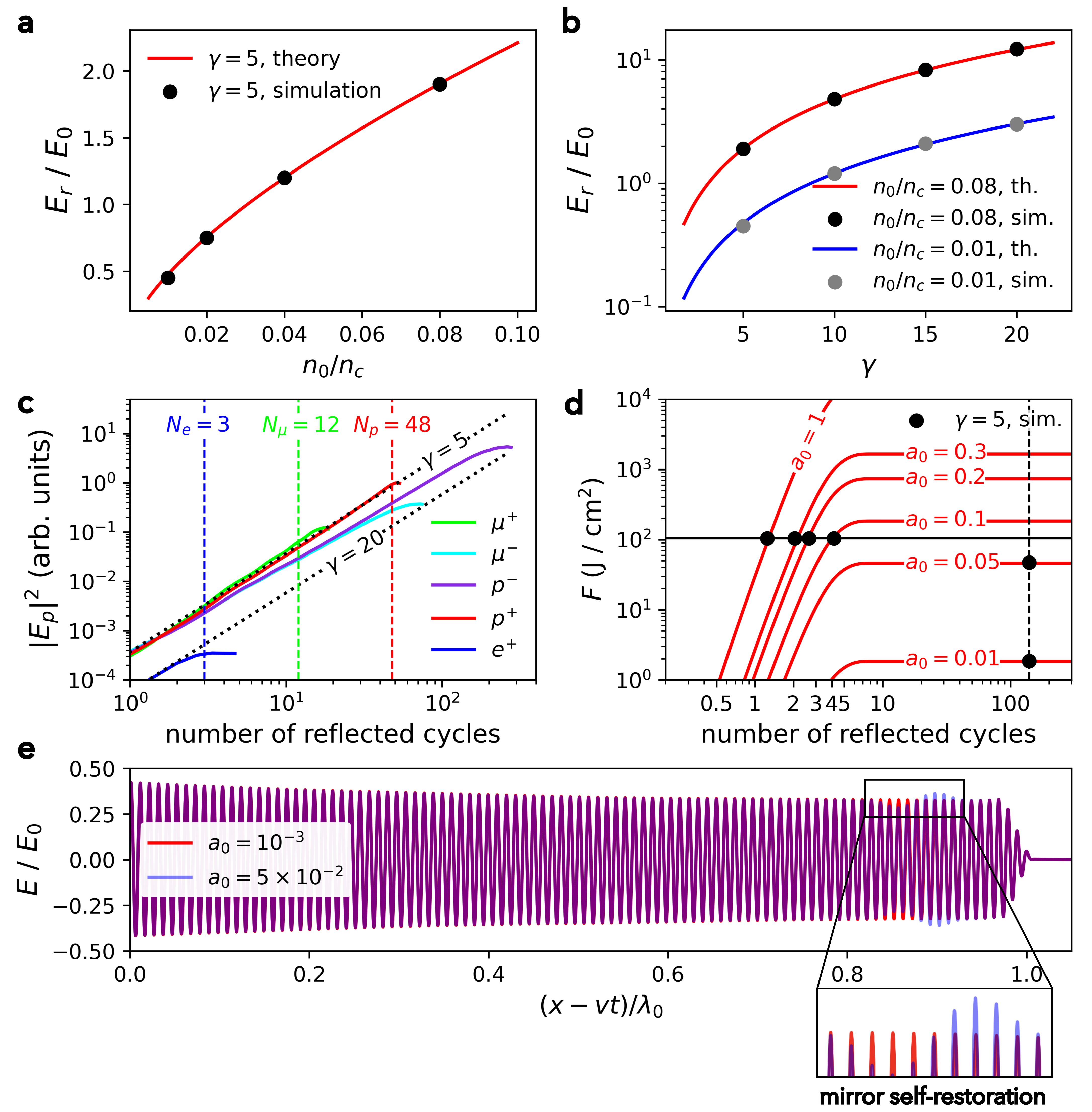}
\caption{\textbf{Properties of beam-driven relativistic mirrors in plasma.} Dependence of reflected amplitude on normalized background plasma density for $\gamma = 5$ (\textbf{a}), and driver Lorentz factor for $n_{0}/n_{c} \in \{0.01, 0.08\}$ (\textbf{b}). Theoretical curves are calculated from Eqs. \ref{eq:1},\ref{eq:7}. \textbf{c}, Evolution of energy spectrum peak $E_{p}$, in terms of number of reflected cycles, for various driver particle types. Theoretical curves are given by $\left|E_{p}\right|^{2} \approx \left|E_{r}\right|^{2}\tau_{r}^{2}$, which is calculated using Eqs. \ref{eq:1}, \ref{eq:7} for $\gamma \in \{5, 20\}$ (dotted). The theoretical number of reflected cycles (vertical, dashed) predicted by Eq. \ref{eq:wb} is calculated as $N \approx 2(L_{wb}-x_{0})/\lambda_{0}$, where $x_{0}$ is the position where the counter-propagating laser initially collides with the mirror. For positrons, muons and protons we have set, respectively, $x_{0}/\lambda_{0} \in \{1.3, 4, 7\}$. Lorentz factor for all particles is set as $\gamma = 5$, except for positrons where $\gamma = 20$. \textbf{d}, Evolution of optical fluence $ F $ incident on a relativistic mirror (red, solid) with $\gamma = 5$, $n_{0}/n_{c} = 0.01$ for various values of normalized laser amplitude. The theoretical value of laser-induced damage threshold (black, solid, horizontal) is given by Eq. \ref{eq:max_fluence} for $n_{0}/n_{c} = 0.01$, and the number of reflected cycles limited due to mirror breaking distance (black,  dashed, vertical) is given by Eq. \ref{eq:wb} as $N \approx 140$ for $x_{0} = 17\lambda_{0}$. \textbf{e}, Temporal profile of coherent X-ray radiation from beam-driven relativistic mirror with $\gamma = 5$, $n_{0}/n_{c} = 0.01$ and $a_{0} \in \{10^{-3}, 0.05\}$. The case of $a_{0} = 0.05$ exhibits the self-restoration property of relativistic mirrors, where the fluence incident on the continuously replenishing nonlinear plasma wave saturates below the damage threshold, if the mirror restoration time $t_{r}$ is short enough, which can be also seen in Fig. 3\textbf{d}.}
\label{fig:3}
\end{figure*}

\subsubsection*{Properties of beam-driven relativistic mirrors and laser-induced damage threshold}

Figs. \ref{fig:3}a,b, show an excellent match between the reflected amplitude given by Eqs. \ref{eq:1},\ref{eq:7}, and the dependence of reflected peak amplitude on normalized background plasma density and mirror velocity obtained from PIC simulations.

\textit{Driver particle type.} To investigate negatively-charged drivers, as well as to verify Eq. \ref{eq:wb}, we again simulated the case with a low-amplitude CW laser with $a_{0} = 10^{-3}$, $n_{0}/n_{c} = 0.08$ and $\gamma = 5$ for protons, muons, their antiparticles, and also the case of $\gamma = 20$ for positrons. Positrons were set faster, because the breaking distance for $\gamma = 5$ is less than the laser wavelength, $L_{wb}\approx 0.7\lambda_{0}$. Similarly, the case of electron drivers is reserved for future study, since the computational requirements necessary to sufficiently resolve the entire elongated nonlinear wake, as well as reflected radiation for $\gamma \geq 20$, are extremely demanding with standard numerical methods.


Fig. \ref{fig:3}c shows the spectrum peak evolution for various drivers. The reflection coefficient for the negatively-charged beams is roughly $70\%$ of the value given by Eq. \ref{eq:7}, but the maximum number of reflected cycles can be up to 5 times larger when comparing antiprotons and protons. As shown in Figs. \ref{fig:1}c,d, this is because stronger longitudinal interior wave modulates the positively-charged drivers as they propagate. Therefore, irrespective of charge number, Eq. \ref{eq:wb} can be interpreted as the minimum mirror propagation distance for a driving beam with particle mass $M$. The number of reflected cycles for the positron case is $N\approx 2.5$, in accordance with Eq. \ref{eq:wb} for $M/m_{e} \approx 1, \gamma = 20$. This highlights the possibility for all-optical beam-driven relativistic mirrors, where positron or electron beams could be produced using compact laser wakefield accelerators \citep{corde2015multi, esarey2009physics, terzani2023compact}.


\textit{Laser-induced damage threshold.} Analogous to standard optical components, relativistic plasma mirrors are susceptible to laser-induced damage \citep{valenta2020recoil}. If the reflected energy is too large, damage can occur due to recoil of the mirror. This recoil threshold can be expressed in terms of maximum fluence on the relativistic mirror, also known as laser-induced damage threshold (LIDT)
\begin{equation}\label{eq:max_fluence}
    \mathrm{LIDT} = \kappa m_e c^2 \frac{n_0 \lambda_{i}}{8|r|^{2}\gamma }.
\end{equation}

Here, $ \kappa $ represents the coefficient of proportionality between the total reflected energy and the total energy of the electrons forming the mirror. Based on simulations with $ a_0 $ ranging from $ 10^{-3} $ to $ 1 $, we deduce that for beam-driven relativistic mirrors $ \kappa \approx 6 \times 10^{-4} $. This yields LIDT $ \approx 100 \ \mathrm{J / cm^2} $ for mirrors with $ \gamma = 5 $, which is two orders of magnitude higher compared to their solid-state counterparts \citep{gallais2014laser, willemsen2022large}. Fig. \ref{fig:3}d compares the LIDT given by Eq. \ref{eq:max_fluence} and the threshold fluence obtained from simulations. Once the LIDT is exceeded, the number of reflected cycles with frequency given by Eq. \ref{eq:1} saturates, but the reflection process does not necessarily cease. However, the properties of reflected radiation may start to differ significantly from those obtained in regime well below the LIDT \citep{valenta2020recoil}.

Furthermore, since beam-driven relativistic mirrors are composed of electrons continuously flowing through, the mirrors possess the ability to self-restore, with a characteristic restoration time of $ t_r \approx \lambda_{i} / c \beta $. The restoration time is particularly short (a few fs) for the mirrors realized by interior waves, given their substantially shorter wavelengths compared to wake waves. Self-restoration must be taken into account when assessing the limits of applicability of Eq. \ref{eq:max_fluence}, i.e., even if the fluence accumulated over a period of time $ \gg t_r $ exceeds the LIDT, the mirror properties remain unaffected, as the impact of laser is fully compensated by the mirror restoration process. The onset of self-restoration is highlighted for the case of $ a_0 = 0.05 $ in Figs. \ref{fig:3}e, where the mirror restoration time is calculated as approximately four laser cycles. The mirror replenishes completely before it is damaged, leading to effective saturation of fluence on the mirror and significantly more reflected cycles than predicted by Eq. \ref{eq:max_fluence}, as shown in Fig. \ref{fig:3}d for $ a_0 \in \{0.01, 0.05\} $. 


To summarize, if the time over which the incident laser deposits damage threshold fluence given by Eq. \ref{eq:max_fluence} is much larger than the mirror restoration time, the mirror will self-restore and continue to produce phase and amplitude stable coherent X-ray radiation. Therefore, large-amplitude few-cycle laser pulses should be used to maximize amplitude of reflected radiation, whereas smaller values of $ a_{0} $ should be considered for a narrowband attosecond X-ray source.  

\subsubsection*{Peak brightness and energy conversion efficiency}

The figure-of-merit for coherent light sources is the peak spectral brightness, $\mathcal{B}$ \citep{attwood_sakdinawat_2017, o2001free}. For coherent X-ray pulses from beam-driven relativistic mirrors, it can be derived as (see Supp. S10 for details)

\begin{figure*}[ht]
\centering
\includegraphics[scale = 0.65]{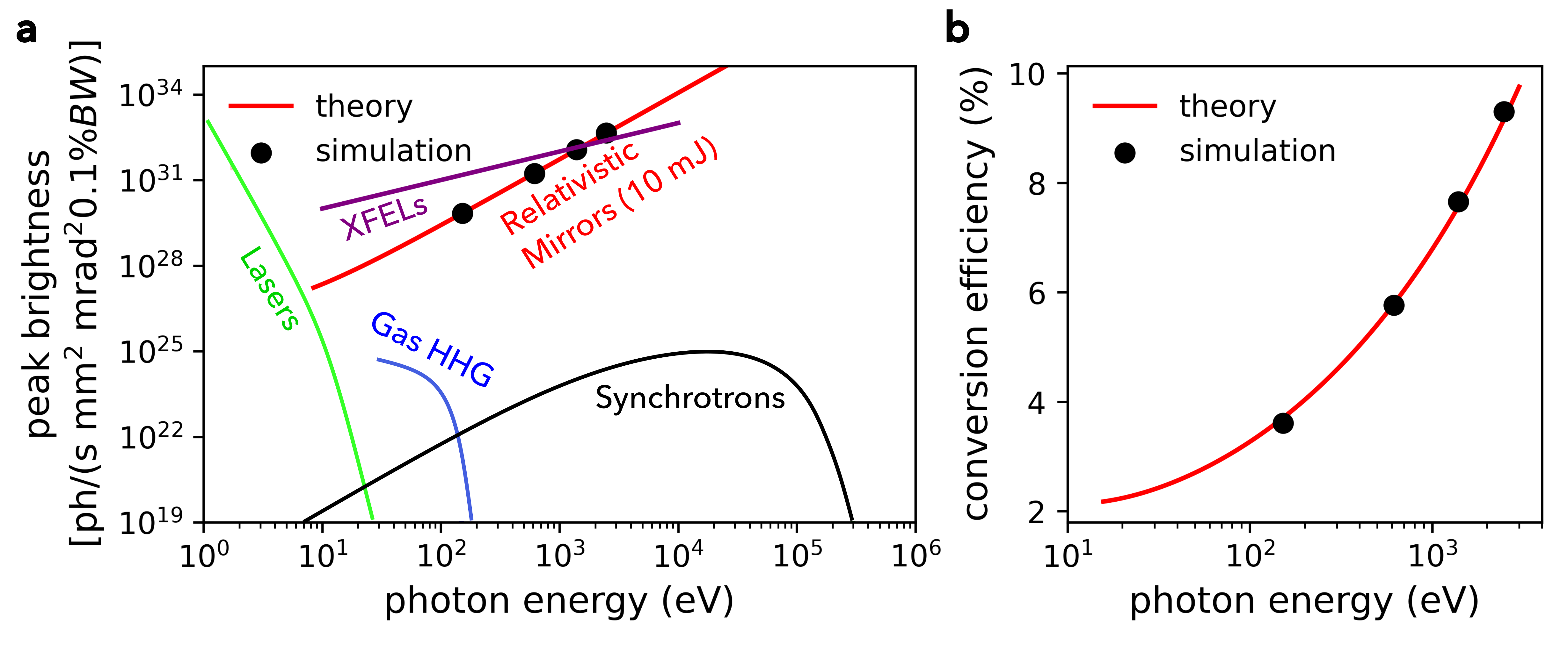}
\caption{\textbf{Properties of coherent X-ray light from beam-driven relativistic mirrors.} \textbf{a}, Peak spectral brightness of $10$ mJ laser pulse reflected from beam-driven relativistic mirrors propagating with $\gamma \in \{5, 10, 15, 20\}$ (black dots) in plasma with normalized electron density $n_{0}/n_{c} = 0.08$. Data for other light sources taken from \citep{boutet2018x}.  \textbf{b}, Incident-to-reflected laser energy conversion efficiency for $n_{0}/n_{c} = 0.08$. Different photon energies correspond respectively to different mirror velocities, $\gamma \in \{5, 10, 15, 20\}$.}
\label{fig:4}
\end{figure*}

\begin{equation}
\begin{split}
\mathcal{B}\left[\frac{\text{photons}}{\text{s mm$^{2}$ mrad$^{2}$ 0.1$\%$ BW}}\right] = &\frac{32c\gamma^{6}N_{\gamma ,i}\left|r\right|^{2}}{10^{15} \lambda_{0}^{3}}\\
 \approx 5\times 10^{30} \gamma^{4}\left(\frac{n_{0}}{n_{c}}\right)^{4/3}&\lambda_{0}^{-4}[\mu\text{m}]\mathcal{E}_{0}[\text{J}],\label{eq:9}
\end{split}
\end{equation}
where $\mathcal{E}_{0}$ is the energy of the incident laser pulse. Fig. \ref{fig:4}a compares Eq. \ref{eq:9} with PIC simulation results and the peak spectral brightness of other currently available coherent light sources \citep{o2001free}. The simulations correspond to background density $n_{0}/n_{c} = 0.08$, proton-driven mirror with $\gamma \in \{5, 10, 15, 20\}$ and incident laser pulse with energy of $10$ mJ. At photon energy of $2.5\text{ keV}$, the peak brightness obtained is $\mathcal{B} \approx 10^{33}$ photons$/($s mm$^{2}$ mrad$^{2}$ 0.1$\%$ BW). Fig. \ref{fig:4}b compares the theoretical conversion efficiency obtained from Eqs. \ref{eq:1},\ref{eq:7} as $\eta \approx 0.36\, (n_{0}/n_{c})^{4/3}\gamma^{2/3}$ to the PIC simulation results. In accordance, conversion efficiency grows with mirror velocity, up to $\eta \approx 9\,\%$ at $2.5\text{ keV}$. 

Beam-driven relativistic mirrors have the potential to be one of the most efficient sources of bright coherent X-ray radiation. Indeed, as long as the maximum fluence on the self-restoring relativistic mirror does not exceed LIDT given by Eq. \ref{eq:max_fluence}, reflected radiation properties are defined by Eqs. \ref{eq:1},\ref{eq:7}, and conversion efficiency increases with reflected photon energy. In this case, beam-driven relativistic mirrors could even realize a novel scheme for laser energy amplification, since $\eta > 1$ for $\gamma > 4.63\,(n_{c}/n_{0})^{2}$. 


\section*{Discussion}

We have presented a new way for generation of bright coherent attosecond X-ray pulses based on laser reflection from relativistic mirrors driven by charged particle beams in plasma. 

In excellent agreement with presented theory, simulation results show that beam-driven relativistic mirrors enable extremely efficient generation of bright, tunable and coherent attosecond X-ray pulses, with simulations showing peak spectral brightness up to $\mathcal{B} \approx 10^{33} \text{ photons/(s mm$^{2}$ mrad$^{2}$ 0.1$\%$ BW)}$ at X-ray photon energy of $2.5$ keV, which is of the order of X-ray free electron lasers. Furthermore, reflectivity of beam-driven relativistic mirrors is robust with respect to mass of the driving particles, differing only in mirror propagation distance, which highlights the possibility for compact all-optical beam-driven relativistic mirrors based on laser wakefield accelerated electron or positron beams. 

The coherent attosecond X-ray pulses are found to be generated over a distance of few micrometers, facilitating a compact source for novel applications where both high intensity and ultrashort pulse pulse duration are required, such as coherent nonlinear X-ray spectroscopy or coherent diffractive imaging of biomolecules. 

Finally, we have found the laser-induced damage threshold of beam-driven relativistic mirrors to be at least two orders of magnitude higher than optical solid-state components. The relativistic mirrors carry a unique property, where the mirror is not damaged unless the threshold fluence accumulates faster than the mirror replenishes itself. In this case, beam-driven relativistic mirrors could be even used as a novel scheme for laser energy amplification.




\section*{Methods}
\subsubsection*{Particle-In-Cell simulations}
\label{s:methods}


We carried out fully-relativistic 1D3V (one spatial and three velocity dimensions) particle-in-cell (PIC) simulations using the code EPOCH \citep{arber2015contemporary}. The parameters of the PIC simulations are defined as follows: The driving particle beams have flat-top profile with length $ L_b $, particle mass $ M/m_{e} \in \{1, 207, 1836\}$, charge density $ \rho $, and the Lorentz factor is varied as $ \gamma \in \{5, 10, 15, 20\} $. 

For positively-charged drivers, $ L_b = \lambda_i $ and $ \rho = 0.96 \, \rho_{wb,+} = 0.96\gamma $, i.e., slightly below the wavebreaking threshold in order to ensure high reflectivity over sufficiently long distance according to Eq. \ref{eq:wb}. For negatively-charged drivers, resolving the full interior plasma wave in addition to the wakefield and the reflected wavelength proved computationally challenging. To reduce the necessary simulation box length required to observe beam-driven wake wave breaking, we shrunk the pulse length and set the particle number density of the driver slightly above the wavebreaking threshold, according to $ \rho = -1/2 < \rho_{wb,-} = -\gamma/(2\gamma + 1) $. The pulse length required to achieve wake wave breaking at $ \gamma = 5 $ was then obtained by numerically solving Eq. \ref{eq:3}, as $ L_{b} \approx 10.7 \, c / \omega_p $. The pulse length was then correspondingly set slightly below this wave breaking threshold, as $ 0.96 \, L_{b} $, for the same reasons as in the case of positively-charged drivers.

The driving particle beam propagates in a pre-ionized homogeneous plasma with number density $ n_0 $, where $ n_0 / n_{c} \in \{0.01, 0.02, 0.04, 0.08\}$. Here, $ n_c $ stands for the critical plasma density corresponding to the wavelength of the counter-propagating laser pulse $ \lambda_0 = 800 \ \mathrm{nm} $. The laser pulse has flat-top profile with smooth $ 4\, T_{0}$ long up-ramp, where $T_{0} = \lambda_{0}/c \approx 2.66\,\mathrm{fs}$ is the laser period, and its normalized amplitude $a_{0}$ is varied from $ 10^{-3} $ to $ 1 $. For the case of a few-cycle Gaussian laser pulse, we have set $a_{0} = 0.1$ for all simulations.

The simulations utilize moving window technique; the window, which moves at the velocity of the driving particle beam, has length of $ 3 \, L_b $. The underlying Cartesian grid is uniform with the resolution of $ 100 $ cells per analytically calculated wavelength of the reflected radiation. The plasma is cold and collisionless, represented with electron quasi-particles moving on the static neutralizing background. Initially, there are $ 10 $ electron quasi-particles per grid cell. The electromagnetic field evolution is calculated using the 2nd order finite-difference time-domain method, whereas the equations of motion for quasi-particles are solved using the Boris algorithm. Absorbing boundary conditions are applied on each of the simulation window boundaries for both the electromagnetic fields and quasi-particles.

\section*{Data Availability}

The data that support the findings of this study are available from the corresponding author upon reasonable request.

\section*{Code Availability}

Numerical PIC simulations were performed with the open source massively parallelized PIC code EPOCH, available at \url{https://github.com/Warwick-Plasma/epoch}. Data analysis and visualization were done using open-source programming language Python and commercial software Wolfram Mathematica. Code used in this study is available from the corresponding author upon reasonable request.

\bibliography{sn-bibliography}

\section*{Acknowledgements}

This work was supported by the project ADONIS (CZ.02.1.01/0.0/0.0/16\_019/0000789) from European Regional Development Fund. This work was supported by the Ministry of Education, Youth and Sports of the Czech Republic through the e-INFRA CZ (ID:90254).

\section*{Author Information}
\subsubsection*{Affiliations}

\noindent \textbf{ELI Beamlines Facility, Extreme Light Infrastructure ERIC, Za Radnicí 835, Dolní Břežany, 25241, Czech Republic}

\vspace{1mm}
\noindent Marcel Lamač, Petr Valenta, Uddhab Chaulagain, Jaroslav Nejdl, Tae-Moon Jeong and Sergei V. Bulanov 
\vspace{2mm}

\noindent \textbf{Faculty of Mathematics and Physics, Charles University, Ke Karlovu 3, Prague 2, 12116, Czech Republic}
\vspace{1mm}

\noindent Marcel Lamač
\vspace{2mm}

\noindent \textbf{Faculty of Nuclear Sciences and Physical Engineering, Czech Technical University in Prague, Břehová 7, Prague 1, 11519, Czech Republic}

\noindent Jaroslav Nejdl
\vspace{2mm}

\noindent \textbf{Kansai Photon Science Institute, National Institutes for Quantum Science and Technology, 8-1-7 Umemidai, Kizugawa, 619-0215, Kyoto, Japan}
\vspace{1mm}

\noindent Sergei V. Bulanov

\subsubsection*{Contributions}

M.L. conceived the research, derived the analytical results and calculated numerical solutions, P.V. performed the particle-in-cell simulations, M.L. and P.V performed analysis of the simulations, processed the data, produced the figures and wrote the manuscript. U.C., J.N., T.M.J. and S.V.B. provided feedback, oversaw and helped to shape the research and the manuscript. All authors reviewed the final manuscript.

\subsubsection*{Corresponding Author}

Correspondence to \href{mailto:marcel.lamac@eli-beams.eu}{Marcel Lamač}.

\section*{Ethics Declaration}
\subsubsection*{Competing Interests}
The authors declare no competing interests.


\end{document}


\title[Bright coherent attosecond X-ray pulses from beam-driven relativistic mirrors (Supplementary Information)\\]{Bright coherent attosecond X-ray pulses from beam-driven relativistic mirrors (Supplementary Information)\\}

\author*[1,2]{\fnm{Marcel} \sur{Lamač}}\email{marcel.lamac@eli-beams.eu}

\author[1]{\fnm{Petr} \sur{Valenta}}

\author[1,3]{\fnm{Jaroslav} \sur{Nejdl}}

\author[1]{\fnm{Uddhab} \sur{Chaulagain}}

\author[1]{\fnm{Tae Moon} \sur{Jeong}}

\author[1,4]{\fnm{Sergey V.} \sur{Bulanov}}

\affil*[1]{\orgdiv{ELI Beamlines Facility}, \orgname{Extreme Light Infrastructure ERIC}, \orgaddress{\street{Za Radnicí 835}, \city{Dolní Břežany}, \postcode{25241}, \country{Czechia}}}

\affil[2]{\orgdiv{Faculty of Mathematics and Physics}, \orgname{Charles University}, \orgaddress{\street{Ke Karlovu 3}, \city{Prague 2}, \postcode{12116}, \country{Czechia}}}

\affil[3]{\orgdiv{Faculty of Nuclear Science and Physical Engineering}, \orgname{Czech Technical University in Prague}, \orgaddress{\street{Břehová 7}, \city{Prague 1}, \postcode{11519}, \country{Czechia}}}

\affil[4]{\orgdiv{Kansai Photon Science Institute}, \orgname{National Institutes for Quantum and Radiological Science and Technology}, \orgaddress{\street{8-1-7 Umemidai, Kizugawa}, \city{Kyoto}, \postcode{619-0215}, \country{Japan}}}

\date{\today}



\maketitle

\newpage

\tableofcontents

\newpage

\section{Theory of weak reflection from relativistic mirrors}
\label{Supp6}

Combining Maxwell's equations and using the Lorenz gauge condition $\partial _{\mu}A^{\mu} = \nabla \cdot \textbf{A} + \partial_{t}\phi /c^{2} = 0$, we obtain the inhomogenous wave equation

\begin{equation}
    \left(\frac{\partial^{2}}{\partial t^{2}} - c^{2}\nabla ^{2}\right)\textbf{A} = \frac{\textbf{j}}{\epsilon_{0}}, \label{wave1}
\end{equation}
where $\textbf{A}$ is the magnetic vector potential, $\epsilon_{0}$ is the vacuum permittivity and $\textbf{j}$ is the current density, acting as a source of radiation. We consider only the one-dimensional geometry of a planar mirror, with a counter-propagating incident laser. Therefore, we have for the transverse derivatives $\partial_{y}\textbf{A} = \partial_{z}\textbf{A} = 0 $. To further simplify the analysis, we consider only linear polarization in the $y$ direction and define the dimensionless vector potential $\textbf{a} = e\textbf{A}/m_{e}c$, reducing Eq. \ref{wave1} into a scalar form. We assume that the radiative currents are due to a propagating electron distribution oscillating in electromagnetic field, therefore $j = - en_{e}v = -ecn_{e}a/\gamma $, where $\gamma$ is the Lorentz factor of the plasma electrons, which gives us the one-dimensional Klein-Gordon equation describing propagation of electromagnetic waves in plasma
\begin{equation}
    \left( \frac{\partial^{2}}{\partial t^{2}} - c^{2}\frac{\partial^{2}}{\partial x^{2}} + \omega_{p}^{2}(x,t)\right) a(x,t) = 0, \label{wave2}    
\end{equation}
where $\omega_{p} = \sqrt{e^{2}n_{e}(x,t)/(m_{e}\epsilon_{0}\gamma)}$ is the relativistic plasma frequency and $n_{e}(x,t)$ is the electron density distribution propagating with normalized velocity $\beta = v/c = \sqrt{1-(1/\gamma)^{2}}$. Counter-propagating radiation scattered from such a relativistic electron distribution is Doppler shifted in frequency, which makes analysis of Eq. \ref{wave2} a non-trivial task in the laboratory frame. We therefore perform a Lorentz boost into the mirror reference frame where both incident and reflected radiation have the same frequency $\omega'$. The $x$ coordinate then transforms as $x' = \gamma(x-vt)$. Since the operator $\Box = (\partial_{t} ^{2} - c^{2}\partial_{x}^{2})$ is Lorentz-invariant, the form of the equation doesn't change and it can be written in the mirror reference frame as
\begin{equation}
    \left(\frac{\partial^{2}}{\partial t'^{2}} - c^{2}\frac{\partial^{2}}{\partial x'^{2}} + \omega_{p}'\text{}^{2}(x')\right) a(x',t') = 0, \label{wave3}    
\end{equation}
where the apostrophe denotes quantities transformed to the mirror rest frame. The normalized vector potential is transformed only in coordinates, because the normalized amplitude ($a_{0} \propto E_{0}/\omega_{0}$) is Lorentz-invariant.

\begin{figure*}
\centering
\includegraphics[scale = 0.4]{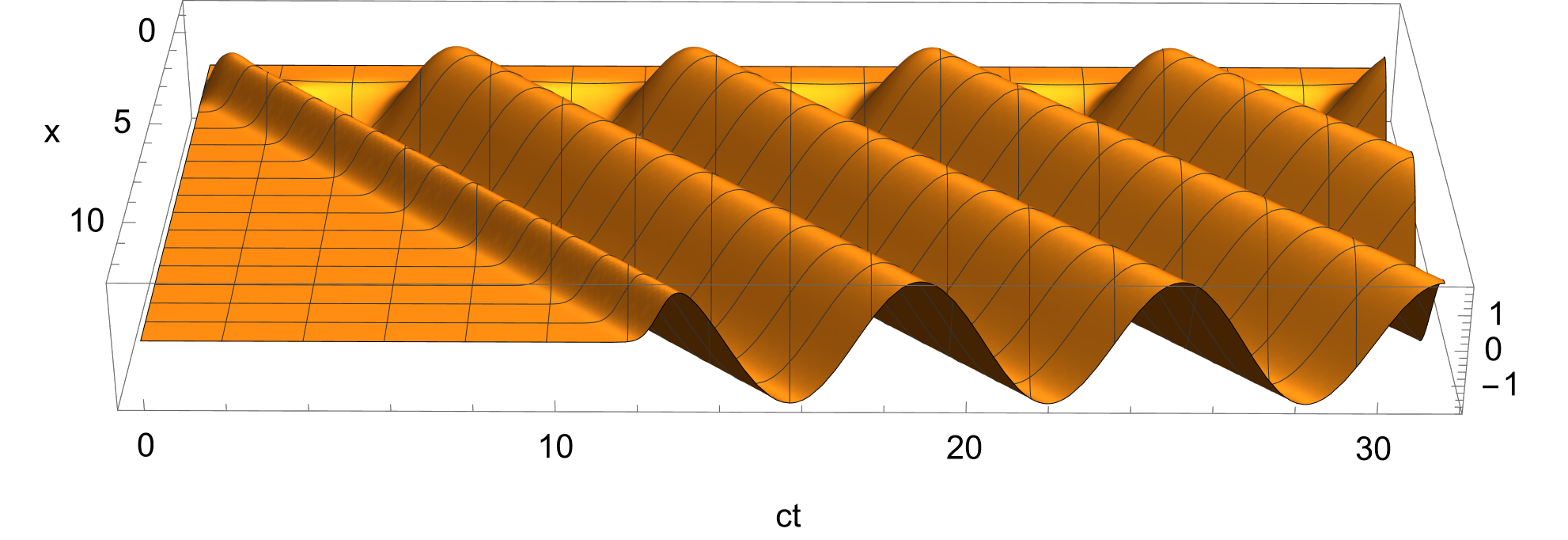}
\caption{Analytical solution of Eq. \ref{wave9} for $k_{p}^{2}(x) = e^{-2x^{2}}$ showing reflected radiation with light rays of constant phase propagating along the light-cone variables $x - ct = const$.}
\label{fig:2}
\end{figure*}

For clarity of the following rest frame analysis, we drop apostrophes for all the transformed variables and return them when necessary. The solution to Eq. \ref{wave3} can be written generally as a superposition of the incident and reflected waves
\begin{equation}
    a(x,t) = a_{0}e^{-i(\omega t+kx-\frac{\pi}{2})} + a_{r}(x,t), \label{wave4}
\end{equation}
where $\omega$ is the radiation frequency and $k = \omega/c$ is the wave number, both evaluated in the mirror rest frame. We include the phase factor $\pi /2$ so that the real part of the incident wave corresponds to a sine wave, such that the real part of the incident wave is zero at spacetime origin $x=ct=0$. Plugging (\ref{wave4}) into (\ref{wave3}) while assuming small reflection, $a_{r}(x,t) \ll a_{0} \approx const.$, gives us the following equation for the reflected wave 
\begin{equation}
     \left(\frac{\partial^{2}}{\partial t^{2}} - c^2\frac{\partial^{2}}{\partial x^{2}}\right)a_{r}(x,t) = -i\omega_{p}^{2}(x)a_{0}e^{-i(\omega t + kx)}. \label{wave5}
\end{equation}

We note that the approximation of weak reflection is equivalent to the physically reasonable assumption that the electrons oscillate with the same phase and normalized transverse momentum as the incident wave, $j = -(ecn_{e}a_{0}/\gamma)e^{-i(\omega t + kx - \frac{\pi}{2})}$. The incident wave $a_{0}(x,t)$ then represents a homogeneous solution to the differential equation \ref{wave1}, while the particular solution $a_{r}(x,t)$ is obtained by solving Eq. \ref{wave1} with the approximated current. 

The natural coordinate system for this problem are the light-cone coordinates $\xi = x-ct, \eta = x+ct$, which correspond to a clockwise $\pi /4$ rotation of the spacetime coordinate system. The derivatives transform as $\partial_{t} + c\partial_{x} = 2c\partial_{\eta}$ and $\partial_{t} - c\partial_{x} = -2c\partial_{\xi}$, which using $ \partial_{t}^{2}-c^{2}\partial_{x}^{2} = (\partial_{t} + c\partial_{x})(\partial_{t} - c\partial_{x})$ gives us the reflected wave equation in terms of the light-cone coordinates as
\begin{equation}
     \frac{\partial^{2}a_{r}}{\partial \eta \partial \xi}(\xi,\eta) = \frac{i}{4}k_{p}^{2}\left(\frac{\eta + \xi}{2}\right)a_{0}e^{-ik\eta}, \label{wave6}
\end{equation}
where $k_{p}(x) = \omega_{p}(x)/c$ is the inhomogeneous relativistic plasma wave number corresponding to the electron distribution. Eq. \ref{wave6} can be solved by direct integration and application of boundary conditions. We first apply the boundary condition that the reflection occurs only for $t > 0$, therefore at $t = 0$ we have $a_{r}(x,0) = 0$. This also implies that the solution exists when $x + ct \geq x - ct$ for all $x$. Integrating the equation over $\eta$, where $\eta \geq \xi$, gives us the derivative of the reflected wave as
\begin{equation}
{\frac{\partial a_{r}}{\partial \xi}(\xi,\eta) = \int_{\xi}^{\eta}\frac{i}{4}k_{p}^{2}\left(\frac{\zeta + \xi}{2}\right)a_{0}e^{-ik\zeta}d\zeta.} \label{wave7}
\end{equation}

Note that, at $t = 0$, Eq. \ref{wave7} satisfies $\partial_{\xi}a_{r}(\eta = x,\xi = x) = 0$, i.e. no radiation is initially emitted along the light cone coordinate $\xi$. To clearly show that Eq. \ref{wave8} describes the reflected radiation, we substitute $\zeta + \xi = 2s$ in the integrand, giving us 
\begin{equation}
\frac{\partial a_{r}}{\partial \xi}(\xi,\eta) = a_{0}\frac{ie^{ik\xi}}{2}\int_{\xi}^{\frac{\eta+\xi}{2}}k_{p}^{2}\left(s\right)e^{-i2ks}ds. \label{wave8}
\end{equation}

If we assume the radiation condition $\partial_{\xi}a_{r} \approx ika_{r}$, we obtain the spatiotemporal profile of the reflected radiation in spacetime coordinates of the mirror rest frame as
\begin{equation}
{a_{r}(x,t) = a_{0}e^{-i(\omega t - kx)}\frac{1}{2k}\int_{x - ct}^{x}k_{p}^{2}\left(s\right)e^{-i2ks}ds}. \label{wave9}
\end{equation}

Eq. \ref{wave9} satisfies the radiation boundary condition $\partial_{\xi}a_{r} = ika_{r}$ exactly for $t\rightarrow \infty $, since then $(\eta + \xi)/2 = x$ and $\xi = x - ct \rightarrow -\infty$, and only the plane wave phase keeps the $\xi$ dependence. This is equivalent to the assumption that, compared to the phase, the amplitude of the wave described by the integral term varies slowly with $\xi$. A more precise solution near could be obtained near the spacetime origin by direct integration of Eq. \ref{wave7}, which leads to a more complicated form. Since we are mostly interested in the far-field behavior, we only keep the term given by Eq. \ref{wave9} which well describes radiation behavior for large $t$ and $x < ct$. 

Finally, the spatiotemporal profile of the reflected wave is obtained by taking the real part of Eq. \ref{wave9}, which is shown in Fig. 2 where a sine wave is reflecting from a Gaussian electron profile. We see that the error of our far-field approximation is small and leads to a small phase shift and reflected amplitude reduction near the line $x = ct$, while the behavior at $x < ct$ shows propagation of reflected radiation with amplitude corresponding to the exact limit value. Precise integration without any approximations of Eq. \ref{wave7} would give us correct phase behavior near the spacetime origin. The absolute value of Eq. \ref{wave9}, corresponding to the amplitude of the reflected wave, reduces in the limit $x \rightarrow \infty$ to the Fourier transform of electron distribution in the mirror rest frame, which was obtained in previous work on relativistic mirrors \citep{bulanov2013relativistic}, and the complex amplitude reflection coefficient $r = a_{r}/a_{0}$ in the far-field limit can be therefore expressed as
\begin{equation}
    r = \frac{1}{2k}\int_{-\infty}^{\infty}k_{p}^{2}\left(s\right)e^{-i2ks}ds, \label{wave10}
\end{equation}
where $k = \sqrt{\frac{1+\beta}{1-\beta}}k_{0}$ is the wave number in the mirror rest frame. The reflected amplitude in the laboratory frame is then simply $E_{r}/(rE_{0}) = (1+\beta)/(1-\beta)$, where $E_{0}$ is the electric field amplitude of the incident laser in the laboratory frame. 

In the following sections, we first analyze and identify properties of beam-driven relativistic mirrors to consequently obtain the electron density distribution required to calculate Eqs. \ref{wave9}-\ref{wave10}.

\section{Beam-driven nonlinear plasma wave equation}
\label{Supp1}

We proceed to analyze plasma waves driven by a charged particle beam in one-dimensional approximation, which is valid for broad drivers satisfying $k_{p}r_{b} \gg 1$, where $k_{p} = \omega_{p}/c$ is the linear plasma wave number, $\omega_{p} = \sqrt{e^{2}n_{0}/\epsilon_{0}m_{e}}$ is the linear electron plasma frequency, $n_{0}$ is the electron plasma density and $r_{b}$ is the radial size of the beam. The Poisson equation and the continuity equation for plasma electrons are then given as
\begin{eqnarray}
&& \frac{\partial ^{2}\phi}{\partial x^{2}} = \frac{e}{\epsilon_{0}}(n_{e} - n_{0} - qn_{b}),\; \label{eq:1}
\\
&& \frac{\partial j_{x}}{\partial x} - e\frac{\partial n_{e}}{\partial t} = 0,\label{eq:2}
\end{eqnarray}
where $n_{b}$ is the number density of the driving beam and $q$ is the charge number of the driving particles, e.g. it is positive for positrons and protons ($q=1$) or ions ($q\geq1$) and it is negative for electrons ($q=-1$). We assume that the driver density satisfies $n_{b}(x,t) = n_{b}(x-vt)$, which means that it propagates in the $x$-direction and its shape does not evolve in time, an assumption that is valid over a relatively long time for ultra-relativistic beams with $\gamma \gg 1$, where $\gamma = 1/\sqrt{1-(v/c)^{2}}$ is the relativistic Lorentz factor of the driving beam and $c$ is the speed of light in vacuum. We now assume that all the plasma quantities also depend only on the Lagrangian coordinate of the driving beam $\xi = x - vt$. This gives us the following transformations for the spatial and temporal derivatives, respectively $\partial _{x} f(\xi (x,t)) = (\partial_{\xi}/\partial_{x})\partial_{\xi} f(\xi) = \partial_{\xi} f(\xi)$ and $\partial _{t} f(\xi (x,t)) = (\partial_{\xi}/\partial_{t})\partial_{\xi} f(\xi) =  -v\partial_{\xi} f(\xi)$. Using these transformations and introducing the following normalized units $\omega_{p}\xi/c = \xi,\text{  } e\phi/m_{e}c^{2} \rightarrow \phi,\text{  } eE/m_{e}c\omega_{p} \rightarrow E,\text{  } n_{e}/n_{0} \rightarrow n_{e},\text{  } qn_{b}/n_{0} = \rho,\text{  } v/c = \beta, \text{  } j/en_{0}c \rightarrow j, \text{  } p/m_{e}c \rightarrow p,$ equations \ref{eq:1}-\ref{eq:2} become
\begin{eqnarray}
&& \frac{\partial ^{2}\phi}{\partial \xi^{2}} = (n_{e} - 1 - \rho),\; \label{eq:3}
\\
&& \frac{\partial}{\partial \xi}\left(n_{e}\beta - n_{e}\beta_{e}\right) = 0,\label{eq:4}
\end{eqnarray}
where $\beta _{e}$ is the normalized velocity of the plasma electrons. Integrating Eq. \ref{eq:4} with the initial conditions $n_{e}(\xi = 0) = 1\text{ and }\beta (\xi = 0) = 0$, which are satisfied by plasma initially at rest, we obtain $n_{e} = 1/(1-\beta_{e}/\beta)$, which shows that extremely large density waves are produced when $\beta_{e}\rightarrow\beta$, and Eq. \ref{eq:3} becomes
\begin{equation}
    \frac{\partial^{2}\phi}{\partial \xi ^{2}} = \frac{\beta}{\beta-\beta_{e}} - 1 - \rho. \label{eq:5}
\end{equation}

It can be shown that the motion of an electron in a one-dimensional nonlinear plasma wave is given by the normalized one-dimensional Hamiltonian, $H(\xi,p_{x}) = \gamma - \beta p_{x} - \phi(\xi)$. For plasma electrons which are initially at rest, the following quantity is conserved, $\gamma - \beta p_{x} = 1 + \phi $. Since only longitudinal fields are considered, the Lorentz factor is simply $\gamma = \sqrt{1 + p_{x}^{2}}$ and Eq. \ref{eq:5} can be cast as
\begin{equation}
    \frac{\partial^{2}\phi}{\partial \xi ^{2}} = \gamma^{2}\left[\beta\left(1 - \frac{1}{\gamma^{2}(1+\phi)^{2}}\right)^{-1/2} - 1\right] - \rho,. \label{eq:6}
\end{equation}

\begin{figure}
\centering
\includegraphics[scale = 0.5]{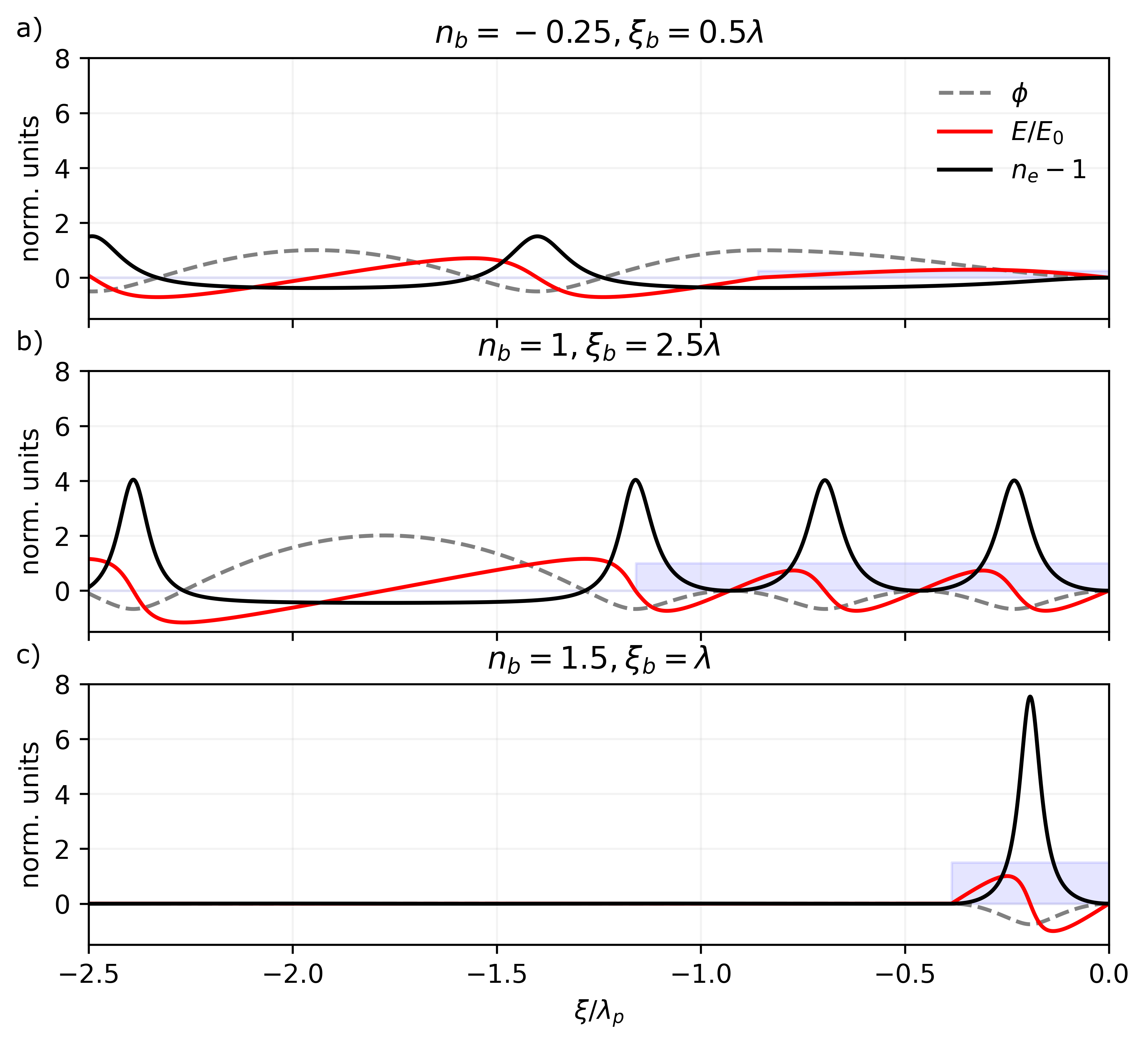}
\caption{\textbf{Poisson equation (\ref{eq:6}) numerical solutions in the limit $\gamma^{2} \gg \rho^{2}$.} Wakefield-optimal solutions with bunch length given by Eq. \ref{eq:15} for (a) $\rho = -0.25$ and (b) $\rho = 1$. Single-cycle interior-wave-optimal solution with bunch length with given by Eq. \ref{eq:14} for $\rho = 1.5$.
\label{fig:1}}
\end{figure}

To obtain analytical results, we approximate the driver as a flat-top beam, $\rho(\xi) = \rho = const.$ for $-L_{b} < \xi < 0$ and zero otherwise, where $L_{b}$ is the normalized driver length. We now proceed to analyze the plasma waves in the region of the ultra-relativistic particle beam, where $-L_{b} < \xi < 0$. Multiplying Eq. \ref{eq:6} by $\phi'$ and integrating with the initial conditions $\phi(\xi = 0) = \phi'(\xi = 0) = 0$, we get the following equation for the electric field of the plasma wave within the driving particle beam,
\begin{equation}
\begin{split}
    \left(\frac{\partial \phi}{\partial \xi}\right)^{2} =& -2(\rho + \gamma^{2})\phi \\ +& 2\beta\gamma^{2}\left(\sqrt{(1+\phi)^{2}-\frac{1}{\gamma^{2}}} - \beta\right). \label{eq:7}
\end{split}
\end{equation}

The existence of plasma oscillations requires the existence of electric field stationary points, which is stated through Eq. \ref{eq:6} as $\phi'' = 0$. From this condition we obtain that real solutions exist only if $\rho > -1/(1+\beta) \approx -1/2$, which tells us that for dense, negatively-charged drivers with $\rho \ll -1/2$, the background plasma cannot support oscillations in the region of the driver due to the strong electrostatic field of the charged beam. For $\rho > -1/(1+\beta)$, equations \ref{eq:6}-\ref{eq:7} allow us to find the amplitude of the nonlinear plasma wave which exists in the interior region of the driver, which we call interior plasma wave from now on. From the electric field null-point condition $\phi' = 0$, we obtain the solutions $\phi _{0} = 0$ and $\phi_{min} = -2\rho(1-1/\gamma^{2})/(2\rho+1+\rho^{2}/\gamma^{2})$, where the zero and non-zero solutions correspond respectively to the interior electron density wave trough and crest, respectively, in the case when $\rho > 0$ and vice versa for $\rho < 0$. 

\section{Lorentz factor of plasma electrons and wave breaking threshold}
\label{Supp2}

Using the conserved Hamiltonian of plasma electrons initially at rest, the electron Lorentz factor can be expressed explicitly in terms of the normalized potential as $\gamma_{e} = \gamma^{2}(1+\phi)\left[1-\beta\sqrt{1-1/(\gamma(1+\phi))^{2}}\right]$, which gives us the Lorentz factor of the electrons in the density peak (corresponding to $\phi_{min}$) of the nonlinear interior wave as
\begin{equation}
    \gamma _{e} = 1 + \frac{2\rho^{2}\beta^{2}}{2\rho+1+(\rho/\gamma)^{2}}.\label{eq:7a}
\end{equation}

The wavebreaking limit of interior plasma waves is given by $\gamma_{e} \rightarrow \gamma$. Solving Eq. \ref{eq:7a} in this limit gives us the following normalized driver density for which wavebreaking occurs immediately upon beam entry into the plasma,
\begin{equation}
  \rho_{wb, \pm}=\begin{cases}
    \gamma, & \text{if $q>0$}.\\
    -\frac{\gamma}{2\gamma+1}, & \text{if $q<0$}.
  \end{cases}\label{eq:7b}
\end{equation}

The driver density corresponding to wavebreaking ($\ref{eq:7b}$) has two solutions, which correspond to positively and negatively charged drivers, respectively.

\section{Wavelength of interior plasma waves}
\label{Supp3}

Inverting Eq. \ref{eq:7} and integrating from $\xi(\phi) < 0$ to $\xi = 0$, we get an explicit dependence of the normalized Lagrangian coordinate $\xi$ on the normalized potential and driver velocity as
\begin{equation}
\begin{split}
    \xi (\phi, \gamma) & = \int_{\phi}^{0}\left[\vphantom{\int_1^2} -2(\rho + \gamma^{2})\phi \right. \\ & \left. + 2\beta\gamma^{2}\left(\sqrt{(1+\phi)^{2}-\frac{1}{\gamma^{2}}} - \beta\right) \vphantom{\int_1^2} \right]^{-1/2} d\phi. \label{eq:10}
\end{split}
\end{equation}

Finally, to obtain the wavelength of the interior plasma waves, we integrate Eq. \ref{eq:10} between the two null-points of the electric field given by $\phi_{0} = 0$ and $\phi_{min} = -2\rho(1-1/\gamma^{2})/(2\rho+1+\rho^{2}/\gamma^{2})$, which gives us the half-period of the interior wave. The wavelength of the nonlinear interior wave normalized to the linear plasma wavelength is then $\lambda_{i} /\lambda_{p} = -2\xi(\phi_{min})/2\pi$. Therefore
\begin{equation}
\begin{split}
    \frac{\lambda_{i}}{\lambda_{p}}(\gamma, \rho) & = \frac{1}{\pi}\int_{\frac{-2\rho(1-1/\gamma^{2})}{(2\rho+1+\rho^{2}/\gamma^{2})}}^{0}\left[\vphantom{\int_1^2} -2(\rho + \gamma^{2})\phi' \right. \\ & \left. + 2\beta\gamma^{2}\left(\sqrt{(1+\phi')^{2}-\frac{1}{\gamma^{2}}} - \beta\right) \vphantom{\int_1^2} \right]^{-1/2} d\phi', \label{eq:10b}
\end{split}
\end{equation}
which is valid when the condition for interior plasma wave existence is satisfied, $\rho > -1/(1+\beta)$. Evaluating the integral \ref{eq:10b} in the limit $\gamma \gg 1$, we get the following expression for the nonlinear interior plasma wave wavelength 
\begin{equation}
    \frac{\lambda_{i}}{\lambda_{p}} = \frac{2}{\pi}\frac{E\left(\frac{2\rho}{2\rho+1}\right)}{\sqrt{2\rho+1}}, \label{eq:11}
\end{equation}
where $E(k)$ is the complete elliptic integral of the second kind. For negatively-charged drivers with $-1/2 < \rho \leq 0$, the plasma wavelength grows with increasing negative charge density and diverges at $\rho \rightarrow -1/2$. For positively-charged drivers with $\rho \gg 1$, expansion of Eq. \ref{eq:11} yields $\lambda_{i} /\lambda_{p} \approx (2/\rho)^{1/2}/\pi$, which reveals a unique mechanism of relativistic nonlinear plasma waves with extremely short-wavelengths, driven by positively-charged particle beams. 

At the wavebreaking limit $\rho \rightarrow \gamma \gg 1$, the wavelength of the nonlinear wave slightly decreases from Eq. \ref{eq:11}, which can be estimated from Eq. \ref{eq:10} in this limit as $\pi\lambda_{i}/\lambda_{p} \approx \left(2/\gamma\right)^{1/2}$. This stands in stark contrast to plasma wakefields driven by either negatively-charged beams or laser pulses, in which case the plasma wave elongates with increasing driver velocity as $\pi\lambda_{i}/\lambda_{p} \approx \left(2\gamma\right)^{1/2}$ \citep{esarey2009physics, rosenzweig1987nonlinear}. 

\section{Optimal driver length}
\label{Supp4}

Now we proceed to analyze the plasma waves behind the driver, where $\xi < -L_{b}$. The Poisson equation (\ref{eq:6}) with $\rho = 0$ is satisfied with initial conditions given by $\phi(-L_{b}), \phi'(-L_{b})$, which correspond to the solutions obtained in the driver region $-L_{b} \leq \xi \leq 0$ due to the continuity of the electric field. Integrating Eq. $\ref{eq:6}$ for $\rho = 0$ with non-zero initial conditions, Eq. \ref{eq:7} takes the following form \citep{akhiezer1956theory}
\begin{equation}
\gamma_{e} + \frac{1}{2}\left(\frac{E_{wake}}{E_{0}}\right)^{2} = C_{0}, \label{eq:13}
\end{equation}
where $C_{0} = \gamma_{e}(-L_{b})+ (-\phi'(-L_{b}))^{2}/2$ is a constant given by initial conditions $\phi(-L_{b}), \phi'(-L_{b})$ at the driver rear $\xi = -L_{b}$. 

When the driver length is exactly the length of a single cycle of the nonlinear plasma wave, $L_{b} = \lambda_{i}$, the potential after a full-cycle gives, as shown above, the initial conditions as $\phi (-L_{b}) = \phi'(-L_{b}) = 0$, and we have $C_{0} = 1$ and the maximum amplitude of the wake is $E_{wake}/E_{0} = \sqrt{2(C_{0}-\gamma_{e})} = 0$, since $1 \leq \gamma_{e} \leq C_{0} = 1$. Therefore, a single-cycle plasma wave is driven optimally and no plasma oscillations occur outside of the region occupied by the ultra-relativistic particle beam. The same analysis holds for optimal multi-cycle waves, in such a case the initial conditions for the Poisson equation are the same and the driver length condition for optimal multi-cycle nonlinear plasma wave generation becomes 
\begin{equation}
    L_{b} = n\lambda_{i}, \label{eq:14}
\end{equation}
where $n$ is a natural number and $\lambda $ is given by Eq. \ref{eq:11}. Any deviation from Eq. \ref{eq:14} gives $C_{0} > 1$ and therefore non-zero plasma wakefield amplitude. The wakefield amplitude is maximized when $\phi'(-L_{b})=0$ and $C_{0} = \gamma_{e}$, where $\gamma_{e}$ is given by Eq. \ref{eq:7a}. These boundary conditions correspond to a driver whose length $L_{b}$ ends exactly at the density peak of the interior plasma wave, as shown in Figs. \ref{fig:1}a,b. The strongest wakefield is therefore created when the driver length satisfies the following resonance condition,
\begin{equation}
    L_{b} = \left(n-\frac{1}{2}\right)\lambda_{i}, \label{eq:15}
\end{equation}
i.e. for the single-cycle wave $m=1$, the driver length fits the first half-cycle of the interior plasma wave, $L_{b} = \lambda_{i}/2$, and the plasma wakefield amplitude is maximized for given driver density and velocity, as shown in Fig. \ref{fig:1}a.

\section{Mirror propagation distance due to relativistic beam-plasma instability}

As a relativistic charged particle beam propagates through plasma, an instability develops which modulates its initially uniform charge density. This modulation eventually breaks the coherent nonlinear oscillations of plasma electrons which sustain the relativistic mirror. To calculate the distance in plasma a beam-driven relativistic mirror propagates before it breaks, we need to find the instability growth rate. 

Let us consider a relativistic charged particle beam with normalized charge density $\rho_{b} = qn_{b}/n_{0}$, mass $M$ and velocity $v$ propagating along the positive direction of the x-axis in plasma composed of electrons and ions with particle number density $n_{0}$. However, since the beam is relativistic, and its longitudinal size is of the order of the electron plasma wave $L_{b}\propto \lambda_{p}$, the influence of slow ions is negligible compared to the fast plasma electrons. The continuity equation and the Euler momentum equation are then given for both relativistic  particle beam and plasma electron species as
\begin{eqnarray}
&& \frac{\partial n_{i}}{\partial t} + \frac{\partial (n_{i}v_{i})}{\partial x} = 0, \label{eq:wb1}
\\
&& m\frac{\partial(\gamma_{i} v_{i})}{\partial t} + m v_{i}\frac{\partial (\gamma_{i}v_{i})}{\partial x} = qE_{x},\label{eq:wb2}
\end{eqnarray}
where $i = b, e$ is the species index and the plasma electrons are initially at rest, $\gamma_{e} \approx 1$. Now, we consider the small-amplitude modes which perturb the the relativistic beam, as $n_{i} \approx n_{0} + \delta n_{i0}$ and $v_{i} \approx v_{i0} + \delta v_{i}$, where $\delta n_{i}, \delta E_{x}, \delta v_{i} \propto e^{i(kx - \omega t)}$. Expanding the momentum in terms of the small perturbation, we get $p_{i} = \gamma_{i}v_{i} \approx \gamma_{i0}v_{i0} + \gamma_{i0}^{3}\delta v_{i}$. Combining the linearized Eqs. \ref{eq:wb1},\ref{eq:wb2} with the Poisson equation, $\epsilon_{0}ik\delta E_{x} = \delta \rho_{b} + \delta \rho_{e}$, we get the following dispersion relation for the relativistic two-stream instability
\begin{equation}
\frac{\omega_{b}^{2}}{\gamma_{b}^{3}(\omega - kv_{b})^{2}} + \frac{\omega_{p}^{2}}{\omega^{2}}= 1,\label{eq:wb3}
\end{equation}
where $\omega_{p}^{2} = e^{2}n_{0}/(m_{e}\epsilon_{0})$ and $\omega_{b}^{2} = (qe)^{2}n_{b}/(M\epsilon_{0}) = (q\rho_{b}m/M)\omega_{p}^{2}$. The $\gamma_{b}^{3}$ factor in the dispersion relation can be intuitively understood as the well-known concept of the longitudinal mass of a particle oscillating along the direction of its relativistic motion, $M \rightarrow \gamma_{b}^{3}M$.
Eq. \ref{eq:wb3} can be rewritten as
\begin{equation}
    \frac{1}{\omega - kv_{b}} = \pm \frac{\gamma_{b}^{3/2}}{\omega_{b}}\sqrt{1-\left(\frac{\omega_{p}}{\omega}\right)^{2}}.\label{eq:wb4}
\end{equation}

Using $\omega = \omega_{r} + i\Gamma$, where $\omega_{r} = \mathcal{R}(\omega)$, and expanding by $\omega_{r} - kv_{b} - i\Gamma $, we get
\begin{equation}
    \frac{\omega_{r} - kv_{b} - i\Gamma}{\omega_{r}^{2} - 2\omega_{r}kv_{b} + (kv_{b})^{2} + \Gamma^{2}} = \pm \frac{\gamma_{b}^{3/2}}{\omega_{b}}\sqrt{1-\left(\frac{\omega_{p}}{\omega}\right)^{2}}.\label{eq:wb5}
\end{equation}

\begin{figure}
\centering
\includegraphics[scale = 0.45]{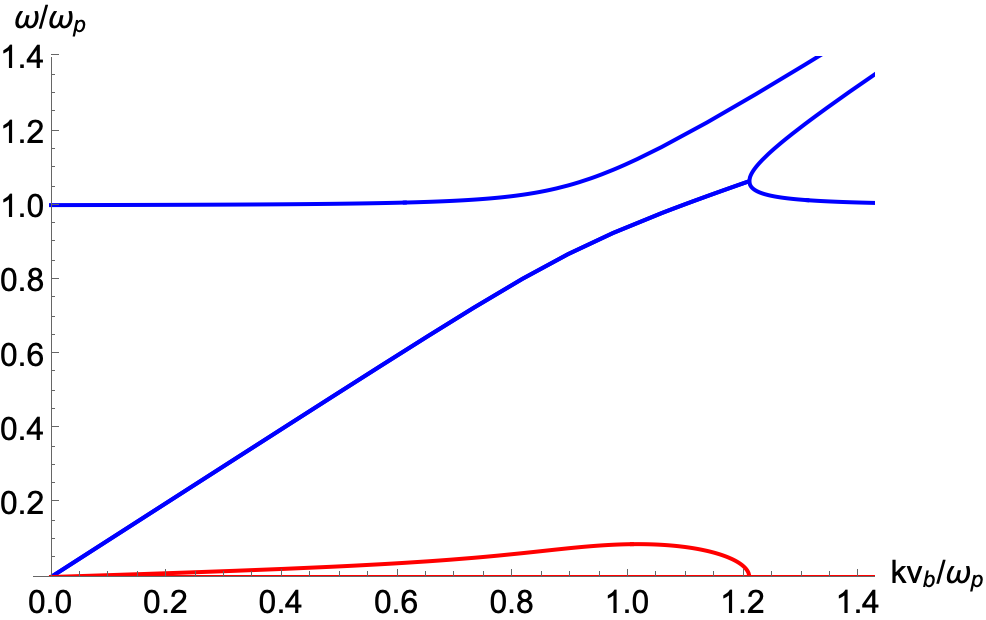}
\caption{Real (blue) and imaginary (red) parts of solution of Eq. \ref{eq:wb3}, calculcated for a positively-charged relativistic particle beam with $qn_{b}/n_{0} = \rho_{wb,+} = \gamma$, $M = m_{e}$ and $\gamma = 20$ (positron). The real part corresponding to the growing unstable modes is given by linearly growing branch, $\omega_{r}\approx kv_{b} \approx kc$.} 
\label{fig:4}
\end{figure}

For even moderately relativistic beams satisfying $v_{b}\approx c$, unstable copropagating electrostatic modes must be also relativistic, with constant phase velocity given by $\omega_{r} \approx ck$. Subsequently, the growth rate must be very small, $\omega = \omega_{r} + i\Gamma \approx \omega_{r} < \omega_{p}$, which can be verified by solving Eq. \ref{eq:wb3} numerically, as shown in Fig. \ref{fig:4}. Using this in Eq. \ref{eq:wb5}, we get the following estimate for the growth rate of the relativistic non-resonant two-stream unstable modes
\begin{equation}
    \frac{\omega_{p}}{\Gamma} \approx  \gamma_{b}^{3/2}\left(\frac{M}{qm_{e}\rho_{b}}\right)^{1/2}\frac{\omega_{p}}{kv_{b}}.\label{eq:wb6}
\end{equation}

Finally, the unstable spectrum observed in PIC simulations is broadband. Typically, the modulation occurs only within a very small region surrounding the copropagating electron density peak of the nonlinear wave (Fig. \ref{fig:1}c), where the electrons pile-up and modulate the driving particle beam. This makes single mode interpretation complicated.

To proceed, we make a generalizing heuristic estimate. Requiring that the mirror breaks immediately, $\tau _{wb}\rightarrow 0$, when $Q \geq Q_{wb}$, where $Q$ the total charge of the relativistic beam and $Q_{wb}$ is the total charge required to produce wave breaking, we get the following estimate that correctly reproduces the timescale ($\tau _{wb} \propto 1/\Gamma$) of the unstable broadband spectra observed in the PIC simulations, written as $\omega_{p}/kv_{b} \rightarrow \sqrt{1 - Q/Q_{wb}}$. This yields the following expression for the beam-driven relativistic mirror propagation distance
\begin{equation}
    k_{p}L_{wb}\approx  N_{e}\gamma_{b}^{3/2}\left(\frac{M}{qm_{e}\rho_{b}}\right)^{1/2}\left(1-\frac{Q}{Q_{wb}}\right)^{1/2},\label{eq:wb7}
\end{equation}
where the factor $N_{e}$ corresponds to the number of e-folds observed in PIC simulations before the mirror breaks. Typically, $N_{e}\approx 1$. For positively-charged drivers with $\rho_{b}\lesssim \rho_{wb,+}\approx \gamma$, we have $Q_{b}/Q_{wb}\approx\rho_{b}/\rho_{wb,+}$ and the mirror propagation distance, for $N_{e} \approx 2\pi/5$, takes the form presented in the main text
\begin{equation}
    \frac{L_{wb,+}}{\lambda_{p}}\approx  \frac{\gamma_{b}}{5}\left(\frac{M}{qm_{e}}\right)^{1/2}\left(1-\frac{\rho_{b}}{\rho_{wb}}\right)^{1/2}.\label{eq:wb8}
\end{equation}

For negatively-charged drivers with $\rho_{b} \approx \rho_{wb,-} = -1/(2+(1/\gamma))\approx -1/2$ and $Q_{b}/Q_{wb} \approx L_{b}/L_{wb}<1$, the mirror propagation distance can be estimated as
\begin{equation}
    \frac{L_{wb,-}}{\lambda_{p}}\approx  \frac{\gamma_{b}^{3/2}}{5}\left(\frac{2M}{|q|m_{e}}\right)^{1/2}\left(1-\frac{L_{b}}{L_{wb}}\right)^{1/2}.\label{eq:wb8}
\end{equation}

\section{Structure of beam-driven interior plasma wave breaking}
\label{Supp5}

As we have shown in S\ref{Supp6}, evaluating reflection of radiation requires electron plasma density distribution of the breaking wave. To gain a foothold, we return to the Poisson equation (\ref{eq:5}) to analyze the structure of interior plasma waves at the wavebreaking threshold, $\gamma_{e}\rightarrow \gamma$. We use the integral of motion $\gamma_{e} - \beta p_{x} - 1 = \phi$ and the identity $1/(1-\beta_{e}/\beta) - 1 = p_{x}/(\beta\gamma_{e} - p_{x})$ to rewrite Eq. \ref{eq:5} as
\begin{equation}
    (\gamma_{e} - \beta p_{x})'' = \frac{p_{x}}{\beta \gamma_{e} - p_{x}} - \rho. \label{eq:19}
\end{equation}

We now expand the momentum around the peak of the nonlinear plasma wave, $p = p_{m} + \delta p + \mathcal{O}(\delta p^{2})$, where $p_{m} = \sqrt{\gamma_{e}^{2}-1}$, where $\gamma_{e}$ is given by Eq. \ref{eq:7a}. At the wavebreaking threshold we have $\gamma_{e}\rightarrow \gamma$, therefore we proceed to use the following power series expansion for left-hand side of Eq. \ref{eq:19}, $\gamma - \beta p_{x} = \sqrt{1+p_{x}^{2}} - \beta p_{x} \approx 1/\gamma + \delta p^{2}/(2\gamma^{3}) + \mathcal{O}(\delta p^{3})$. For the right-hand side of Eq. \ref{eq:19} we use $\gamma_{e} \rightarrow \gamma$ and therefore $\beta\gamma - p_{x} = \beta\sqrt{1+p_{x}^{2}} - p_{x} \approx -\delta p/\gamma^{2} + \beta\delta p^{2}/(2\gamma^{3}) + \mathcal{O}(\delta p^{3})$. Keeping only the leading order terms in Eq. \ref{eq:19}, we get
\begin{equation}
    (\delta p^{2})'' = -2\gamma^{6}\frac{\beta }{\delta p} - \gamma. \label{eq:20}
\end{equation}

Multiplying Eq. \ref{eq:20} by $(\delta p^{2})'$ and integrating over $\xi$, we get
\begin{equation}
    (\delta p'\delta p)^{2} + 2\gamma^{6}\beta \delta p + \gamma\delta p^{2}/2 = C, \label{eq:21}
\end{equation}
where $C$ is an integration constant. In the limit $\delta p \rightarrow 0$, the solution of Eq. \ref{eq:21} is given by the value of the integration constant $C$. If $\delta p' \delta p \rightarrow 0$ when $\delta p \rightarrow 0$, then we have $C = 0$ and up to the leading order Eq. \ref{eq:21} takes the following form
\begin{equation}
    (\delta p')^{2} = -2\gamma^{6}\frac{\beta}{\delta p}. \label{eq:22}
\end{equation}
In the limit $\xi \rightarrow 0$, where $\xi$ is the distance from the position of the density peak, a solution of Eq. \ref{eq:21} is given as
\begin{equation}
   \delta p = -\left(\frac{3}{\sqrt{2}} \gamma^{3}\beta^{1/2}\xi \right)^{2/3}.\label{eq:23}
\end{equation}

The velocity can be expanded up to the first order as $\beta_{e} \approx \beta + \frac{\delta p}{\gamma^{3}} + \mathcal{O}(\delta p^{2})$, therefore $\beta_{e} \approx \beta -\frac{1}{\gamma}\left(\frac{3}{\sqrt{2}} \beta^{1/2}\xi \right)^{2/3}$ and the normalized electron density becomes
\begin{equation}
   \frac{n_{e}}{n_{0}} = \frac{\beta}{\beta-\beta_{e}} \approx \gamma\left(\frac{\sqrt{2}}{3}\frac{\beta}{\xi}\right)^{2/3}.\label{eq:24}
\end{equation}

Eq. \ref{eq:24} shows that the density structure of a breaking interior plasma wave has the form of a cusp caustic \citep{arnold2003catastrophe}, which has an integrable singularity at $\xi \rightarrow 0$, therefore the total number of electrons remains finite even for the asymptotic leading-order term given by Eq. \ref{eq:24}. It was previously shown that such a cusp singularity is a general feature of electron plasma wave breaking \citep{panchenko2008interaction, bulanov2013relativistic, bulanov1998particle}. The structure of beam-driven wave breaking is not different, since the effect of driver density represented by the third term of Eq. \ref{eq:21} only contributes in the higher order corrections.

\section{Reflection coefficient of beam-driven relativistic mirror at the wave breaking threshold}

\label{Supp7}

In the laboratory frame, the electron distribution given by the asymptotic cusp distribution (\ref{eq:24}) can be written as 

\begin{equation}
\frac{n_{e}}{n_{0}} = \gamma\left(\frac{\sqrt{2}}{3}\frac{\beta}{k_{p}(x-vt)}\right)^{2/3}. \label{rc1}
\end{equation} 

\begin{figure}
\centering
\includegraphics[scale = 0.4]{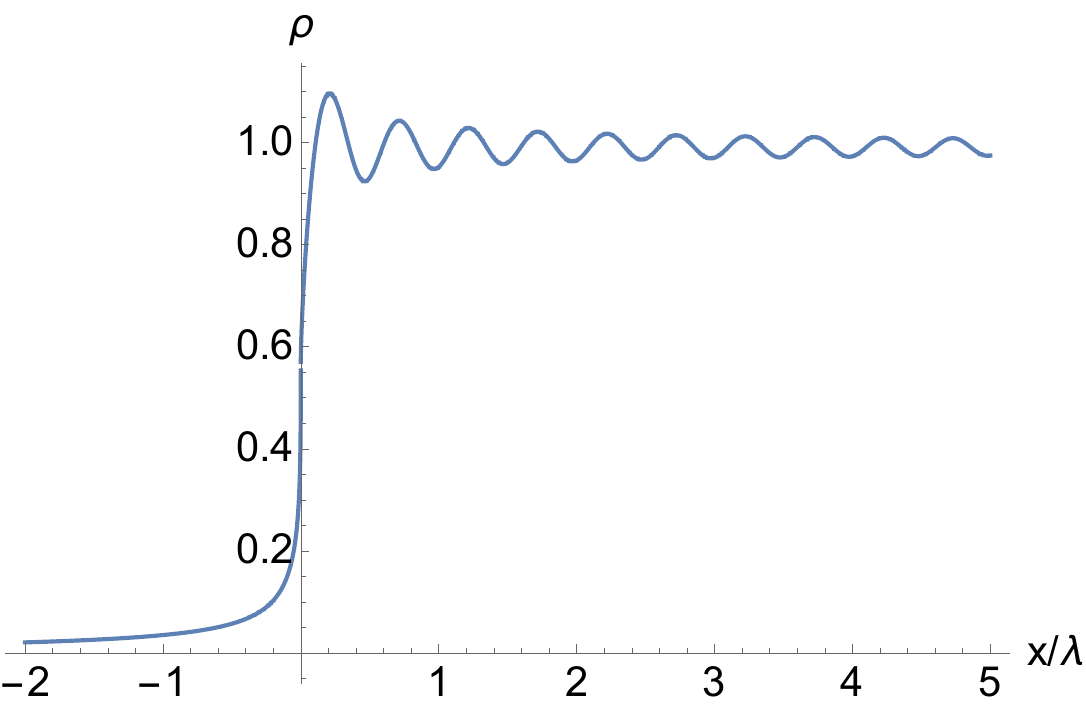}
\caption{Spatial profile of complex amplitude reflection coefficient near the origin of the mirror rest frame ($x=0$). Here $\lambda$ represents the rest frame radiation wavelength. The reflected wave amplitude is modulated at the second harmonic of the incident radiation near the mirror origin. In the limit $x\rightarrow \infty$, the amplitude stabilizes at the value given by Eq. \ref{rc3}.} 
\label{fig:3}
\end{figure}

We now transforming into the mirror rest frame using $x' = \gamma (x-vt)$. The electron charge density is part of the four-current which satisfies the continuity equation $\partial_{\mu}j^{\mu} = 0$, and it must therefore transform as $\rho' = \gamma(\rho - j_{x}\beta/c)$ due to charge conservation. Transforming the spacetime coordinates, the rest frame electron density distribution then becomes $n_{e}'/n_{0}' = \gamma(\sqrt{2}\beta \gamma /3)^{2/3}(x')\text{}^{-2/3}$, where $n_{0}' = n_{0}/\gamma $ is the background plasma density in the mirror rest frame. We therefore have the following inhomogenous relativistic plasma wave number (using $\beta \approx 1$)
\begin{equation}
k_{p}'^{2}(x') = \left(\frac{2}{9}\right)^{1/3}\frac{k_{p}^{4/3}\gamma^{2/3}}{x'^{2/3}}, \label{rc2}
\end{equation} 

where $k_{p} = \omega_{p0}/c$ is the linear plasma wave number calculated in the laboratory frame. Using Eq. \ref{rc2} to calculate the amplitude reflection coefficient given by Eq. \ref{wave10}, we get
\begin{equation}
    r = \kappa \left(\frac{n_{0}}{\gamma n_{c}}\right)^{2/3},\label{rc3}
\end{equation}
where $\kappa = \pi/(2^{4/3} 3^{2/3}\Gamma \left(\frac{2}{3}\right))\approx 0.44$ and $\Gamma(z)$ is the Euler gamma function. Complex amplitude reflection coefficient (\ref{rc3}) was previously obtained in the context of laser-driven relativistic mirrors \citep{bulanov2013relativistic, panchenko2008interaction}. Calculating Eq. \ref{wave9} in the limit $t\rightarrow \infty$, we can further obtain the spatial profile of the reflected amplitude around the mirror position, which is shown in Fig. \ref{fig:3}.

In PIC simulations, we observe that Eq. \ref{rc3} overestimates amplitude of the reflected radiation. Indeed, Eq. \ref{rc3} is derived using the singular cusp distribution. However, in fully kinetic simulations electron density is finite both before, during and after wavebreaking, which is the expected behavior due to the discrete number of electrons (and macroparticles in the case of particle-in-cell simulations), which compose nonlinear plasma waves. 

To amend this, we introduce the finite distribution scale length $l$, also known as sharpness parameter \citep{bulanov2013relativistic}, and generalize the electron distribution in the laboratory frame, given by Eq. \ref{rc1}, as

\begin{equation}
\frac{n_{e}}{n_{0}} = \frac{\gamma 2^{1/3}}{3^{2/3}k_{p}^{2/3}}\left(\frac{\beta^{2}}{l^{2} + (x-vt)^{2}}\right)^{1/3}, \label{rc4}
\end{equation} 
where the limit $l\rightarrow 0$ reproduces the dominant asymptotic cusp structure obtained by solving Eq. \ref{eq:19}. Transforming Eq. \ref{rc4} into the rest frame, we get the following rest frame inhomogeneous plasma wave number
\begin{equation}
k_{p}'^{2}(x') = \left(\frac{2}{9}\right)^{1/3}\frac{k_{p}^{4/3}\gamma^{2/3}}{l'^{2}+x'^{2}}, \label{rc4a}
\end{equation} 
where $l' = \gamma l$ is the finite distribution length scale in the mirror rest frame. The rest frame reflection coefficient evaluated in terms of laboratory frame parameters then becomes
\begin{equation}
    r \approx \frac{\pi^{1/2}}{2^{5/6} 3^{2/3}\Gamma \left(\frac{1}{3}\right)}\frac{k_{p}^{4/3}l^{1/6}}{\gamma^{1/3}k_{0}^{7/6}}K_{1/6}\left(4\gamma ^{2} k_{0}l\right),\label{rc5}
\end{equation}
where $K_{n}(x)$ is the modified Bessel function of the second kind. By measuring the finite distribution scale length $l$ from simulations, the reflection coefficient is precisely given by Eq. \ref{rc5}. Fully-coherent emission will occur when a distribution of radiating particles satisfies $k_{r}l\ll 1$, where $k_{r}$ is the wave number of the reflected radiation in the laboratory frame, $k_{r} \approx 4\gamma^{2}k_{0}$. 

In the PIC simulations, we observe finite distribution length scales of the order $k_{r}l_{sim}\approx 3.5\times10^{-2}$, therefore we estimate $4\gamma^{2}k_{0}l \approx 1/(9\pi)$ and the rest frame amplitude reflection coefficient can be written in the following simple form
\begin{equation}
    r \approx \alpha \left(\frac{n_{0}}{n_{c}\gamma}\right)^{2/3},\label{rc6}
\end{equation}
where $ \alpha = \frac{\pi^{1/3}}{2^{1/6}3^{2}\Gamma(1/3)}K_{1/6}\left(\frac{1}{3^{2}\pi}\right) \approx 0.3$ is the proportionality constant obtained by introducing finite distribution length scale $l$. The scaling follows that of the cusp reflection coefficient given by Eq. \ref{rc3}, but the coefficient is roughly $30\%$ smaller due to the electron density finiteness. As shown with the results presented in the main text, Eq. \ref{rc6} perfectly describes the particle-in-cell simulation results of relativistic mirrors close to the wave breaking threshold in the parameter range $\gamma \in \{5,10,15,20\}$ and $n_{0}/n_{c} \in \{0.01, 0.02, 0.04, 0.08\}$. 

\section{Energy spectrum of reflected coherent radiation}
\label{Supp8}

If we assume that the radiation is reflected with constant amplitude reflection coefficient, the electric field in the spectral domain of a signal with duration $\tau_{r}$ can be expressed in the laboratory frame as
\begin{eqnarray}
\begin{split}
    E_{r}(x,\omega) =&  E_{r}\int_{-\infty}^{\infty} e^{-i(\omega_{r}t-k_{r}x)}e^{i\omega t}dt \\ =& E_{r} e^{ik_{r}x} \int_{0}^{\tau_{r}}e^{i(\omega -\omega_{r})t}dt, \label{specin1} 
\end{split}
\end{eqnarray}
where $E_{r} = rE_{0}(1+\beta)/(1-\beta)$ is the reflected amplitude and $E_{0}$ is the incident electric field amplitude. Therefore, after directly evaluating the last integral in Eq. \ref{specin1}, the normalized energy spectrum can be expressed as
\begin{equation}
    \left|\frac{E(\omega)}{E_{p}}\right|^{2} = \text{sinc}^{2}\left(\frac{(\omega - \omega_{r})\tau_{r}}{2}\right), \label{specin2}
\end{equation}
where $\text{sinc}(x) = \text{sin}(x)/x$ is the cardinal sine function, $E_{p}^{2} = \left|E_{r}\right|^{2}\tau_{r}^{2}$ is the value of the spectrum peak at $\omega = \omega_{r}$, which grows quadratically in time due to temporal coherence, and $\tau_{r}$, $\omega_{r}$ are respectively the double Doppler shifted reflected angular frequency and pulse duration.

\section{Peak spectral brightness}
\label{Supp9}

The value of peak spectral brightness for a monochromatic light source reflected from a homogeneous mirror in terms of the conventional unit of photons per second per relative spectral bandwidth of $10^{-3}$ per transverse radiation phase space can be written as 

\begin{equation}
    \mathcal{B} = \frac{N_{ph,r}(N/1000)}{2\pi^{2}\theta^{2}\sigma^{2}\tau_{r}},
\end{equation}

where $N$ is the number of reflected cycles corresponding to the relative spectral bandwidth as $\Delta \omega / \omega = 1/N$, $N_{ph,r}$ is the total number of reflected photons, $\theta$ is the half-angle divergence of reflected radiation and $\rho$ is the waist of the incident laser. The reflected divergence is $\theta \approx \theta_{0}/(2\gamma)^{2}$, where $\theta_{0}$ is the half-angle divergence of the incident laser. If we consider a diffraction limited incident laser and a homogeneous mirror, the beam parameter product characterizing transverse phase space of incident radiation becomes $\rho \theta_{0} = \lambda_{0}/\pi$, where $\lambda_{0}$ is the incident laser wavelength. The peak spectral brightness is then
\begin{equation}
    \mathcal{B} = \frac{8\gamma^{4}N_{ph,r} (N/1000)}{\lambda_{0}^{2}\tau_{r}}.
\end{equation}

Finally, considering the reflected pulse duration satisfies $\tau_{r} \approx \tau_{0}/(2\gamma)^{2} \approx (N\lambda_{0}/c)/(2\gamma)^{2}$, we obtain the final formula for the peak spectral brightness of radiation reflected from a homogeneous beam-driven relativistic mirror,
\begin{equation}
    \mathcal{B} = \frac{32c\gamma^{6}N_{ph,r}}{1000 \lambda_{0}^{3}}.
\end{equation}

The ratio of reflected to incident number of photons can be calculated using the rest frame reflection coefficient as $N_{ph,r}/N_{ph,i} = \abs{r}^{2}$. Finally, to convert to the conventional synchrotron unit of $\text{photons/(s mm}^{2}\text{ mrad}^{2}\text{ }0.1\%\text{BW)}$, we use m$^{2}$rad$^{2} = 10^{12}$mm$^{2}$mrad$^{2}$ to obtain
\begin{equation}
\mathcal{B}\left[\frac{\text{photons}}{\text{s mm$^{2}$ mrad$^{2}$ 0.1$\%$ BW}}\right] = \frac{32c\gamma^{6}N_{ph,i}\abs{r}^{2}}{10^{15} \lambda_{0}^{3}},
\end{equation}
where the reflection coefficient is given by Eq. \ref{rc6}.

\pagebreak

\bibliography{sn-bibliography}
